# Zero-threshold optical gain in electrochemically doped nanoplatelets and the physics behind it

Jaco J. Geuchies[†,*], Robbert Dijkhuizen[†], Marijn Koel[†], Gianluca Grimaldi[†,$], Indy du Fossé[†], Wiel H. Evers[†], Zeger Hens[#], Arjan J. Houtepen[†]

[†] Optoelectronic Materials Section, Faculty of Applied Sciences, Delft University of Technology, Van der Maasweg 9, 2629 HAZ Delft, The Netherlands

[#] Department of Chemistry and Center for Nano and Biophotonics, Ghent University, 9000 Ghent, Belgium;

[*] Current address: Max Planck Institute for Polymer Research, 55128 Mainz, Germany.

[$] Current address: Center for Nanophotonics, AMOLF, 1098 XG Amsterdam, the Netherlands / Cavendish Laboratory, University of Cambridge, CB2 1TN Cambridge, United Kingdom.

KEYWORDS (Word Style "BG_Keywords"). Nanoplatelets, femto-second transient absorption spectroscopy, electrochemistry, doping, optical gain.

**ABSTRACT**

Colloidal nanoplatelets (NPLs) are promising materials for lasing applications. The properties are usually discussed in the framework of 2D materials, where strong excitonic effects dominate the optical properties near the band edge. At the same time, NPLs have finite lateral dimensions such that NPLs are not true extended 2D structures. Here we study the photophysics and gain properties of CdSe/CdS/ZnS core-shell-shell NPLs upon electrochemical *n* doping and optical excitation. Steady-state absorption and PL spectroscopy show that excitonic effects are weaker in core-shell-shell nanoplatelets due to the reduced exciton binding energy. Transient absorption studies reveal a gain threshold of only one excitation per nanoplatelet. Using electrochemical *n* doping we observe the complete bleaching of the band edge exciton transitions. Combining electrochemical doping with transient absorption spectroscopy we demonstrate that the gain threshold is fully removed over a broad spectral range and gain coefficients of several thousand cm$^{-1}$ are obtained. These doped NPLs are the best performing colloidal nanomaterial gain medium reported to date. The low exciton binding energy due to the CdS and ZnS shells, in combination with the relatively small lateral size of the NPLs, result in excited states that are effectively delocalised over the entire platelet. Core-shell NPLs are thus on the border between strong confinement in QDs and dominant Coulombic effects in 2D materials. We demonstrate that this limit is in effect ideal for optical gain, and that it results in an optimal lateral size of the platelets where the gain threshold per nm$^2$ is minimal.

# INTRODUCTION

Colloidal semiconductor nanomaterials have the potential to form efficient lasers. Solution processing allows facile integration into various device architectures, including conformal coating on patterned substrates.[1,2] The size tunability due to quantum confinement translates in control over the gain spectrum and lasing color. Most studies in this field have focused on CdSe quantum dots (QDs), where the underlying photophysics are broadly understood.[3] Recently it was shown by various groups, including ours[4–7], that $n$ doping allows to reduce the threshold for optical gain. At the same time, of the various colloidal nanomaterials that have been investigated, 2D nanoplatelets (NPLs) of II-VI semiconductors have stood out with the lowest gain threshold and the highest gain coefficients.[8–11] This provokes the suggestion that $n$-doped nanoplatelets could be the ultimate colloidal gain material. However, substantiating this hypothesis with a theoretical description of gain in doped NPLs is complicated by the fact that the photophysics in 2D NPLs is quite different from that in 0D QDs.

In QDs, the optical properties are dominated by quantum confinement, which is stronger than Coulomb interactions (the so-called "strong confinement regime"). The development of gain is dominated by state filling, and Coulomb interactions between carriers act only as a perturbation. In contrast, the optical properties in NPLs are usually considered to be those of 2D semiconductors, dominated by excitonic effects, due to strong Coulomb interactions between electrons and holes. Optical excitation or charging of NPLs with electrons results in state filling and a bleach of the excitonic and free carrier absorption[12], but in addition results in screening of both the electron-hole Coulomb and exchange interactions by the additional carriers. Such screening is expected to reduce the exciton binding energy, which decreases the oscillator strength of the exciton transition, and leads to lifetime broadening due to increased scattering and increases exchange interactions[13,14]. As the screening increases with exciton or electron density, a Mott transition to an electron-hole plasma may take place.[15]

In a hydrogenic model, the 2D exciton binding energy is four times the bulk (3D) exciton binding energy. This increase is further enhanced by dielectric confinement in colloidal NPLs. As a result, the exciton binding energy in CdSe NPLs is typically ~200 meV (whereas the bulk exciton binding energy is 15 meV)[10] and excitons are extremely robust at room temperature. The high exciton binding energy corresponds to very small exciton Bohr radii[14,16], which in turn means that state filling of the exciton transition is minor, even at high excitation densities.[11] Complete bleaching of the exciton transition has thus not been observed in CdSe NPLs and gain by free carriers or exciton molecules strongly suffers from competing exciton absorption.

Here we investigate the development of gain in optically excited and electrochemically *n* doped CdSe/CdS/ZnS core-shell-shell (CSS) NPLs. The use of core-shell-shell NPLs is necessary for stable and reproducible electrochemical doping, but also results in a much smaller exciton binding energy. The core-shell structure effectively eliminates dielectric confinement, such that the exciton binding energy is of the order of 40 meV. These excitons are much larger than in core only CdSe NPLs, and thus more sensitive to screening and state filling. We demonstrate that we can fully bleach the optical transitions by either photoexcitation or by electrochemical *n* doping. Remarkably, in undoped NPLs, we find a gain threshold of only one excitation per nanoplatelet. We combine *n* doping with photoexcitation in ultrafast spectroelectrochemical measurements to probe the development of optical gain in doped NPLs. We find the lowest gain threshold reported for colloidal nanomaterials, and gain spectra with a broad bandwidth. In addition, the gain coefficients are 3-4 times higher than in the best *n* doped QD films and show no sign of saturation. This demonstrates that doped NPLs are indeed an extremely promising gain medium. The results show that the low exciton binding energy due to the CdS and ZnS shell, in combination with the relatively small lateral size of the NPLs, results in excited states that are effectively delocalised over the entire platelet, which consequently behave as a particle in a box. These core-shell nanoplatelets are on the border between strong confinement in QDs and dominant Coulombic effects resulting in excitonic behaviour in 2D materials. We demonstrate that this limit is in effect ideal for optical gain, and that it results in an optimal lateral size of ~500 nm$^2$ of the platelets where the gain threshold per nm$^2$ is minimal.

The paper is organized as follows: we start by discussing the steady state optical properties of CdSe/CdS/ZnS NPLs of varying shell thickness. We then controllably add excitons, electrons, or both to investigate the effects of state filling and screening. Finally, we quantify optical gain in films of *n* doped NPLs and demonstrate that the gain threshold vanishes at the highest doping density.

**RESULTS AND DISCUSSION**

EXCITONS IN CORE-SHELL(-SHELL) NPLs?

We synthesized zinc-blende CdSe NPLs which emit around 510 nm via procedures outlined in the methods section. We have grown CdS and ZnS layers around the CdSe core NPLs by a continuous injection method[2,17,18], in order to enhance their photoluminescence quantum yield (PLQY) and their photo- and electrochemical[19] stability. In Figure 1(a), we present the absorption spectra for the CdSe/CdS NPLs with shell thicknesses varying from one to six monolayers. This results in a gradual red-shift of the absorption spectrum with increasing CdS shell thickness. Moreover, the intensity ratio between the heavy hole (HH) and light hole (LH) absorption gradually decreases. This is not due to a relative increase in oscillator strength of the LH transition, but rather due to a decrease in exciton binding energy, a consequence of the reduced dielectric contrast when a shell is grown, which results in free carrier absorption that overlaps with the LH line.

Using methods outlined in literature[10,13,20], we decompose the absorption spectra of the core-only and core-shell NPLs in excitonic and free-carrier absorption to get an estimate of the HH-exciton binding energy. The full analysis is shown in the SI, and the results are shown as the inset of Figure 1(a). The HH exciton binding energy ($E_{B,HH}$) of 254 meV in the core-only CdSe NPLs decreases to 89 meV upon growing one monolayer of CdS on the NPL surface. For the core-shell-shell CdSe/CdS/ZnS NPLs used throughout this study, which are coated with six monolayers of CdS and two monolayers of ZnS, we estimate $E_{B,HH}$ to be around 42 meV, which is in between the bulk (15 meV)[21] and the 2D limit (4 times bulk = 60 meV) exciton binding energies[22] (see SI section S2). In the 2D hydrogenic model a binding energy of 42 meV corresponds to a 2D exciton Bohr radius of $a_{B,2D}$ 3.2 nm, compared to $a_{B,2D} = 0.53$ nm for the core only NPLs. The final CdSe/CdS/ZnS NPLs, presented in Figure 1(b), have a PLQY of

62% and have lateral dimensions of 24.8±1.6 nm by 9.9±0.6 nm. Especially the ZnS shell enhances the electrochemical stability, similar to observations we made on CdSe/CdS/ZnS QDs[19], as shown in the SI (section S3).

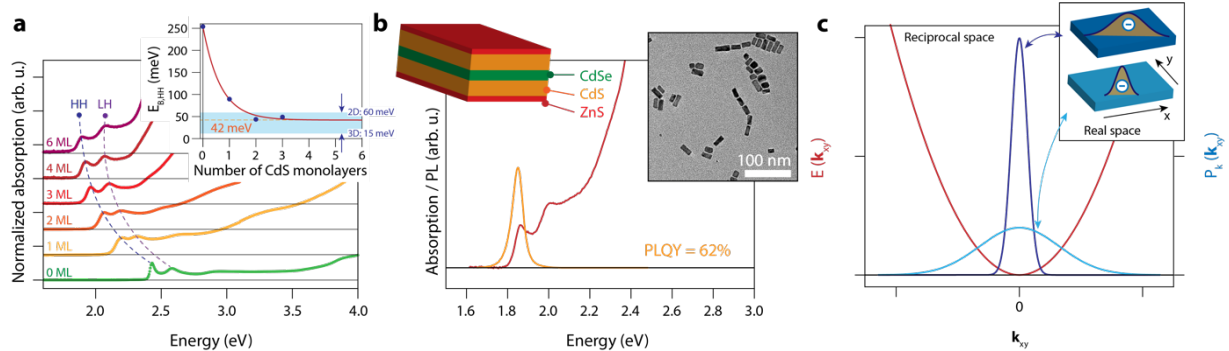

**Figure 1: Steady-state optical properties of CdSe/CdS(/ZnS) core-shell(-shell) nanoplatelets. (a)** Optical absorption of CdSe-based NPLs with increasing shell thickness. The inset shows the estimated HH exciton binding energy as a function of shell-thickness. For the main NPLs studied throughout this work (6 MLs of CdS, 2 MLs of ZnS), we estimate an exciton binding energy of 42 meV. **(b)** Absorption and photoluminescence spectra of a dispersion of CdSe/CdS/ZnS NPLs in hexane. The inset shows a representative TEM micrograph. **(c)** Schematic of the dispersion relation $E(\mathbf{k}_{xy})$ versus in-plane wavevector $\mathbf{k}_{xy}$ and probability amplitude $P_k(\mathbf{k}_{xy})$ of the electron wavefunction in real (inset) and reciprocal space.

To check whether, due to the lower $E_{B,HH}$, excitons still form in these CSS NPLs, we performed temperature dependent absorption and PL measurements down to 13K, presented in the Figure S12. In contrast to more traditional quantum-wells, such as GaAs[23] ($E_{b,HH}$ ≤ 10 meV[24]), there is no distinct sign of a crossover from excitonic to free carrier transitions[25]. This could either mean that $E_{B,HH}$ is still so high that the optical features at room temperature are still dominated by excitons, or it could mean that the electron and hole wavefunctions are fully delocalized over the NPLs and the optical features are the result of quantum confinement, *vide infra*.

## OPTICAL GAIN IN CSS NPLs

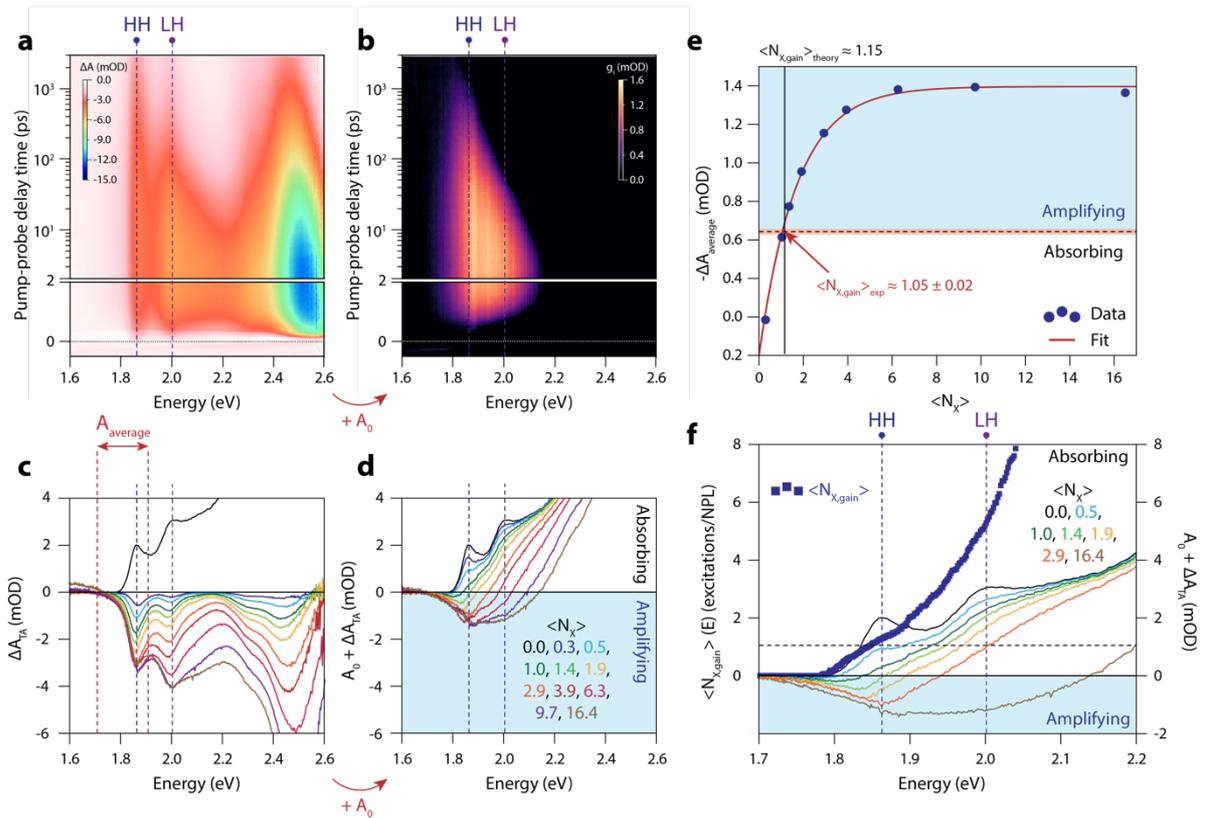

**Figure 2: Optical gain characteristics of isolated CSS NPLs in solution @400 nm pump.** (a) Transient absorption map for $<N_X> = 16.4$. (b) Optical gain map of (a), obtained by adding the steady-state absorption to the TA signal ($A_0 + \Delta A_{TA}$). (c) Spectral slices of the TA map at 5 ps pump-probe delay time, for increasing excitation density. (d) Gain spectra at 5 ps for increasing excitation density. (e) Determination of the spectrally averaged (between 1.7-1.9 eV) optical gain threshold for the HH transition, $<N_{X,gain}> = 1.05 \pm 0.02$ excitations/NPL. (f) Energy-dependent threshold for optical gain, by repeating the analysis shown in (e) for each energy. On the red side of the HH transition, less than one excitation per NPL will lead to optical gain.

Next, we analyze the effect of photoexciting CSS NPLs dispersed in hexane with fs Transient Absorption (TA) spectroscopy. We measure the change in absorption $\Delta A_{TA}$ upon excitation with 400 nm pump-pulses as a function of time-delay and excitation density. Measuring the fluence-dependence of the bleach amplitude allows to obtain the absorption cross-section by assuming Poisson excitation statistics and Auger recombination of multi-excitons.[7] Following this approach we determined the cross-section at 400 nm, $\sigma_{400\,nm}$, to be $3.3 \cdot 10^{-14}$ cm$^2$. As outlined in the SI (section S5), we used multiple alternative procedures to verify this cross section, as it is important to determine the excitation density per NPL $<N_X> = J\sigma_{400\,nm}$, where $J$ is the photon fluence per cm$^2$ per laser pulse.

Figure 2(a) shows a TA map for an excitation density of $<N_X> = 16.4$. The HH and LH bleach maxima are indicated with dashed lines. Figure 2(b) shows an optical gain map, acquired by adding the steady-state absorption $A_0$ to the $\Delta A_{TA}$ signal, to obtain the excited-state absorption A'. Positive absorption is color-coded black, whereas negative absorption, *i.e.* optical gain, is color coded according to the optical density. Note that in the gain map, the HH and LH transitions are strongly broadened, and the lifetime of the gain signal is longest on the red side.

Spectral slices through the TA and optical gain data are presented in Figure 2(c) and (d) respectively, for a pump-probe delay time of 10 ps. While clear bleach features of the HH and LH exciton lines are observed, in the optical gain spectra, there is no sign of gain from the distinct transitions. Instead, it appears that optical gain develops first on the red side at low fluence, consistent with work on core-only NPLs, that attribute this to gain from biexciton molecules[10]. At higher fluence, the gain spectrum broadens, spanning both the HH and LH transition and featuring a spectral width of roughly 500 meV. Strikingly, we observe a complete bleach of the HH and LH exciton absorption features at low excitation densities ($<N_X> \sim 1$), in stark contrast to work on core-only CdSe NPLs[11], where upon creating 112 excitations on average per NPL, the HH and LH transitions were still prominently visible in the excited-state absorption spectrum.

To quantitatively capture the photophysics for optical gain, and determine $<N_{X,\,gain}>$ for the HH transition, we spectrally average the steady-state absorption and bleach amplitude (between the dashed red lines in Figure 2(c)) for each fluence. This spectral averaging corrects for spectral shifts and reveals the total change in the absorption strength of the HH transition, due to state filling and stimulated emission or a reduction in the oscillator strength. As shown in Figure 2(e), we find that $<N_{X,\,gain}> = 1.05 \pm 0.02$, a figure we will come back to shortly, which is equivalent to a 2D exciton density of $4 \cdot 10^{11}$ cm$^{-2}$. We note that this gain threshold is an average; the absorption cross section depends on the orientation of the NPLs and hence there will be a distribution of excitation densities over the orientations in the NPL ensemble. The gain threshold of roughly one per NPL should be interpreted not as exact, but as an indication that only a few excitations per NPL are required to achieve optical gain. Furthermore, we determined the gain threshold for all probe energies, to obtain a so-called gain threshold spectrum $N_{X,gain}(E)$, which is shown as the blue solid squares in Figure 2(f). For energies below the HH transition, we obtain optical gain for vanishingly low excitation densities in a spectral range up to 1.8 eV, where there is significant light amplification. The distinction between spectrally averaged and single wavelength gain thresholds is important, although often neglected in the literature.

To understand what it means that the spectrally averaged gain threshold in these CSS NPLs is only ~1 we consider the effect of state filling on excitons in 2D materials. Such an exciton can be written as a linear combination of free carrier states, as shown schematically in Figure 1(c). To form a small, localized exciton in real space requires the combination of many free carrier states with different momenta. The widths of the wavefunction in real space and momentum space are related via the Heisenberg uncertainty relation. Optical excitation or electronic doping leads to the occupation of some free carrier states that are now no longer available for the formation of new excitons. This causes a decrease of the exciton absorption, an effect that is called state filling. In the case of a large exciton binding energy, and a correspondingly small exciton Bohr radius, the number of k-states that contribute to the exciton wavefunction is large, and the bleach of a single free carrier (occupying a state near k = 0) is small. For a smaller binding energy this bleach will be larger. The low-gain threshold, where only ~one excitation per NPL is required to achieve transparency, demonstrates that only one electron and one hole free-carrier state contribute to the excited state wavefunction, i.e., the excited state wavefunction is effectively delocalized over the entire nanoplatelet. This implies that the Coulomb interaction between electron and hole is not the dominant factor in controlling the wavefunction, but that its spatial extend is given by the size of the NPLs. This can be rationalized by the low exciton binding energy, due to reduced dielectric contrast between the core CdSe and the CdS and ZnS shells compared to oleate ligands, in combination with the relatively small lateral sizes of the NPLs.

Due to the large exciton size, the wavefunction of the excited state will approach that of a particle in a box. If we calculate the sum of the confinement energy of electrons and holes in the lateral dimensions of the NPLs using a simple rectangular particle in a box model, we find a value of 49 meV, see SI section S6 for details, almost the same as the exciton binding energy. This suggests that these NPLs are on the border between weak and strong confinement, i.e. on the border between single carriers and excitons. We can further estimate that the difference in energy of the electrons between the first and $2^{nd}$ quantized level in the longest dimension of the NPLs is ~20 meV. This implies that at room temperature, most of the photoexcited electrons occupy the lowest quantized energy level.

For a truly extended 2D system the lateral size is not relevant for the gain properties. However, the NPLs investigated here are not extended 2D systems. An interesting question is how, in this limit, the lateral size influences the gain properties, and whether an optimum lateral size exists.

This optimum should be the size where the gain threshold density, measured in excitations per nm$^2$, is lowest, as this corresponds to a medium that can show gain at the lowest photon fluence (in Jcm$^{-2}$s$^{-1}$ for continuous wave excitation).

To this end, we theoretically treat the NPL as a particle-in-a-box. We neglect excitonic effects and assume the energy levels of a particle-in-a-box with dimension given by the thickness of the NPLs, as well as the two lateral dimensions. The details are given in the SI section S6. Further, we assume a quasi-Fermi-Dirac distribution in the conduction and valence band and compute the excitation density, in terms of the number of excitations per nm$^2$ that are required to achieve transparency. Figure S14 shows how the threshold density depends on the lateral sizes. A minimum is found for square platelets with lateral sizes of 23 nm, where the threshold density is around 0.012 excitations per nm$^2$. Reducing the lateral area of a NPL increases the threshold excitation density (in excitations/nm$^2$),

For small lateral sizes, a similar number of excitations per NPL (around 1) is required. Hence, reducing the lateral area of the NPLs increases the threshold excitation density in excitations/nm$^2$. For larger lateral sizes, the energy spacing in both the conduction and valence band becomes so small that significant thermal population of higher levels occurs. This leads to an increase of the gain threshold per NPLs. The simulations show that the latter effect becomes dominant, such that an optimum lateral size exists around 23×23 nm$^2$, where the threshold density is minimal. The optimum platelet is square, since the energy spacing scales as the length$^{-2}$ so that for a given area, the thermal population of higher levels is minimal if the two lateral lengths are the same. Importantly this shows that, in absence of excitonic effects, relatively small NPLs are better gain materials than extended 2D systems of the same thickness.

# ELECTROCHEMICAL *N* DOPING OF CSS NPL-SOLIDS

To reduce the gain threshold further, we electrochemically charged the NPLs with electrons. We used spectroelectrochemistry (SEC) to determine the reproducibility and efficiency of electron injection. NPL films on ITO were prepared by consecutive dipcoating steps, as outlined in the SI. Figure 3(a) shows the absorption (top) and photoluminescence (PL, bottom) SEC data as a function of potential between the NPLs-on-ITO working electrode vs. a Ag pseudoreference (PRE) electrode (with a calibrated potential of +4.35V vs vacuum, see SI Figure S11). Upon decreasing the potential, the Fermi level of the ITO working electrode is raised and electrons are injected into the NPL solid. This is clearly demonstrated by the absorption bleach of several transitions, most notably the HH and LH transitions. The reversibility of charging and discharging is shown by sweeping the potential three times between the open circuit voltage ($V_{OC}$ = -0.3V) and -1.5V. Note that core-only NPLs and CdSe/CdS core crown NPLs can also be charged but the maximum achievable doping density (judged from the absorption bleach) was much smaller, and the PL quenches irreversibly (see SI). This is indicative of permanent electrochemical changes on the surface of the NPLs, likely related to the partial reduction of surface Cd ions.[26,27] In contrast, using CdSe/CdS/ZnS NPLs, we can controllably and reproducibly add electrons via electrochemical doping.

In Figure 3(b) we show absorption spectra at different potentials, obtained by adding the change in absorption due to electrochemically injected electrons, $\Delta A_{SEC}$, to the steady-state absorption of the film ($\Delta A_{SEC} + A_0$). We observe that the absorption at the HH transition is fully bleached, whereas at the energy of the LH transition, a broad and featureless absorption band with no signs of distinct excitonic transitions remains. We attribute this residual absorption to free-carrier absorption, whose onset overlaps with the LH transition.

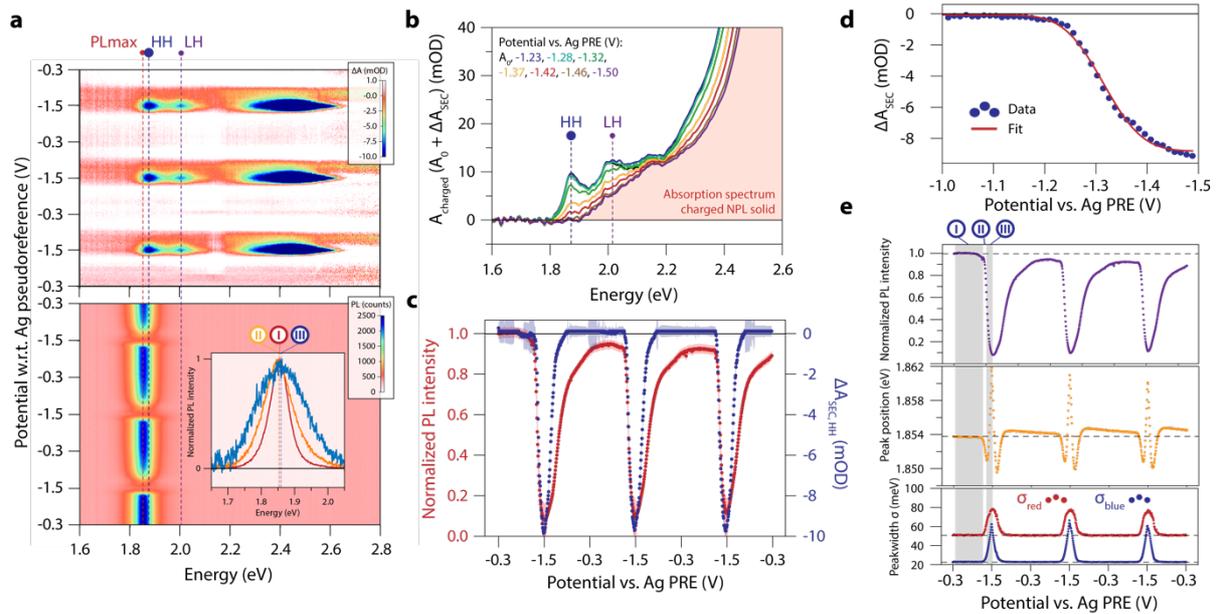

**Figure 3: Spectroelectrochemistry on a CSS NPL film.** (a) Absorption (top) and photoluminescence (bottom) SEC of a CSS NPL film. The inset in the bottom SEC-PL map shows individual PL spectra for the neutral film (-0.3V, red line), at -1.32V (most redshifted, yellow line) and -1.5V (blue line). (b) Absorption spectra at different potentials. The distinct excitonic transitions (HH and LH) collapse and a broad, featureless absorption remains when charging the NPLs at -1.5 V vs. the Ag PRE. (c) Normalized absorption bleach of the HH transition (blue) and PL intensity (red) during 3 potential sweeps. Note that the PL recovery is slower than the quenching; 5 minutes after the experiment, the original PL intensity is fully retrieved. (d) Absorption bleach amplitude vs. electrochemical potential. The red line is a fit to a sigmoidal function. (e) Fitted peak amplitude, peak position and standard deviation (on the red and blue side of the PL maximum) of the PL peak as a function of electrochemical potential.

In Figure 3(c) we compare the changes in the film absorption and PL as a function of the electrochemical potential. At around -1.25V we observe that both the excitonic absorption bleaches and the PL quenches. This shows that electrons are injected into the conduction band; state filling reduces the absorption and Auger recombination reduces the PL efficiency. The fact that the PL and absorption decrease at the same potential is an indication that the films are relatively trap-free and stable, since traps or electrochemical surface reactions often lead to PL changes when the Fermi level is still in the bandgap.[19,28] The PL amplitude drops slightly over the course of consecutive potential cycles but is restored to its initial intensity within five minutes after the end of the experiment, suggesting that a small amount of charge is stored in relatively shallow traps.

To quantify both the absorption and PL SEC data in more detail, we fitted two Gaussian bleach functions to the absorption changes, and an asymmetric Gaussian function (with a different width on the red and blue side) to the PL data; the fitted parameters are presented in Figure 3(d) and (e). The onset of the absorption bleach lies around -1.25V vs. the Ag PRE (corresponding to -3.1 eV below vacuum) and a complete bleach of the exciton absorption is observed around -1.40V. The shape of $\Delta A(V)$ is well-described by an error function with a standard deviation of 60 mV. The PL data show that upon electron injection, the PL energy first redshifts by 3 meV, possibly due to shake-up processes[29], after which it blueshifts by 11 meV, likely to recombination of injected electrons in higher states in the conduction band. Moreover, the PL also broadens significantly. The broadening is stronger on the blue-side of the PL spectrum than on the red-side.

The electrochemical results show that we can completely bleach the HH exciton absorption by injecting electrons into the NPLs. We can estimate to what energy the Fermi level needs to be raised above the band edge (a value we term $\Delta E_{tot}$) to cause a complete bleach of the exciton absorption due to state filling based on a simple Heisenberg model (outlined in the SI, section S7). The rationale is that from the exciton Bohr radius, the electron momentum can be estimated via the Heisenberg relation. When all states up to this momentum, which corresponds to $\Delta E_{tot}$ via the electron dispersion, are occupied, this causes a complete bleach due to state filling. As shown in the SI one can derive that

$$\Delta E_{tot} = \frac{\hbar^2}{4m_e^* m_0 a_{X,2D}^2} \tag{2}$$

If we use the previously estimated 2D Bohr radius of 3.2 nm, we find that $\Delta E_{tot}$ should equal 14 meV, which is much less than the ~120 mV that is needed to bleach the exciton line experimentally (Fig. 3d). This observation suggests that state-filling alone should more quickly bleach the excitonic transitions than we observe. A possible explanation is that the potential change in the electrochemical experiment does not correspond one-to-one to the change of the Fermi level. The applied potential may drop partially over the ITO/NPL interface (as desired) and partially over the NPL/electrolyte interface. The latter part will not lead to a change in the Fermi level. In addition, possible side reactions, such as the reduction of molecular oxygen[30] could reduce the actual charge density, and hence Fermi level. In this case the system is in a steady state rather than in a true equilibrium and the potential difference that is applied is larger than the change in the Fermi level.

ULTRAFAST SPECTROELECTROCHEMISTRY

The TA and the SEC results presented above show that the excitonic absorption can fully bleach upon photoexcitation or electrochemical *n* doping. The resulting optical features appear broad and resemble free carriers. An open question is whether state filling alone is responsible for the disappearance of the exciton absorption, or whether Coulomb screening and enhanced electron-hole exchange interactions in the photoexcited or *n* doped samples are also partially responsible for these observations. To test this, we combined electrochemical doping with fsTA measurement. We consider that if *only* state filling is responsible for the bleach of the excitonic transitions upon *n* doping, photoexcitation could still result in the formation of excitons. Their stimulated emission should show up as narrow features in fsTA measurements on doped films, with a spectral shape similar to the original exciton absorption, as depicted in Figure 4(a). If however, due to screening and increased exchange interactions, only free carriers remain in doped films, such sharp SE signal will not be observed, as depicted in Figure 4(b).

The fsTA experiments are shown in Fig. 4, using a pump wavelength of 400 nm, on a neutral (Fig. 4(c)) and charged (at -1.5V vs. PRE, Fig. 4(d)) NPL film. These TA maps were measured in the linear excitation density regime ($<N_X> = 0.2$). The neutral (c, e) and charged (d, f) film show strong differences in their TA responses. The HH (1.87 eV) and LH (2.0 eV) transitions appear as sharp and narrow bleach features in the neutral film with a lifetime that exceeds the pump-probe time delay window of our setup (3 ns). For the charged film (Fig. 4 d, f) we observe two broadened bleach features near the band-edge; a weak, broad bleach that is redshifted from the neutral HH transition by 90 meV, and a second bleach that has its center at the LH peak position. Compared to the neutral film, where the lifetime of the bleach features extends beyond the time window of our setup (3 ns), the bleach features in the charged film decay within ~500 ps, as a result of efficient Auger recombination in the charged film. Furthermore, the absorption bleach at higher energy (> 2.4 eV) is much weaker, and blueshifted, compared to the neutral film.

At short pump-probe delay times (< 1 ps), we also observe differences between the charged and neutral NPL film. For the neutral film, in addition to an absorption bleach, there is some induced absorption (IA) on the red side of the HH and LH/free carrier transitions. This is often ascribed to band-gap renormalization, and/or hot-carrier effects that result in a spectral shift of the absorption transitions. In contrast, for the charged film, there is no HH absorption in the ground state, so we do not observe a redshift of this feature. In contrast we observe induced absorption at the HH energy, probably as a result of a transient redshift of free carrier absorption. Negative $\Delta A$ of the HH transition appears after 1 ps in the charged film. Negative $\Delta A$ necessarily comes from stimulated emission (since there is no absorption at this energy in the ground state of the charged NPLs) and the onset of SE at the HH energy reflects the cooling of holes from the LH to the HH band.

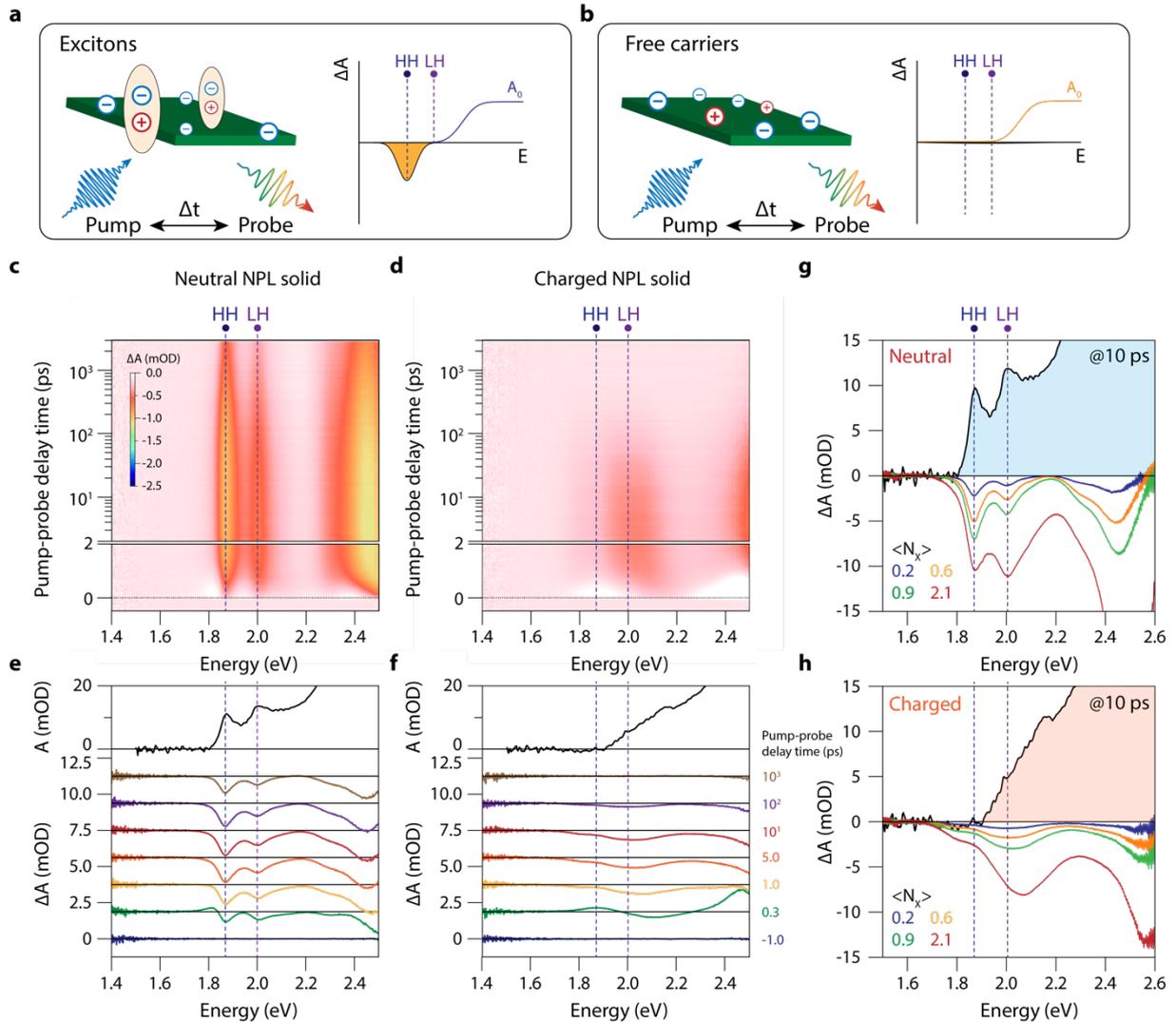

**Figure 4: TA of the neutral and fully charged (-1.5V vs. Ag PRE) film after excitation at 400 nm (3.1 eV).** (a) Schematic of the expected TA spectrum for doped NPLs where excitons are formed. This should result in narrow bleach features, resulting from SE from the excitons. (b) Schematic of the expected TA spectrum for doped NPLs where only free charges form. Here, the absence of SE from excitons should result in the absence of narrow bleach features. (c) TA map of the neutral NPL film, with $<N_X> = 0.2$. A clear bleach of both the HH and LH excitons is observed. (d) TA map of the charged NPL film, with $<N_X> = 0.2$. The distinct excitonic bleaches (HH and LH) disappear and we observe SE from the HH exciton and bleaching/SE of the free-carrier transitions. (e) Spectral slices through the TA data of the neutral NPL film in (c) at different pump-probe delay times. (f) Spectral slices through the TA data of the charged NPL film in **(d)** at different pump-probe delay times. The distinct excitonic bleaches are replaced by broader features. **(g)** $\Delta A_{TA}$ spectra of the neutral NPL solid at 10 ps pump-probe delay time. Upon increasing the excitation density, the HH transition is fully bleached ($\Delta A_{TA} > A_0$), and narrow bleach features of the excitonic transitions are observed. **(h)** $\Delta A_{TA}$ spectra of the charged NPL solid at 10 ps pump-probe delay time. Already at low excitation densities, the spectra look different from the neutral NPL solid; the features are shifted, broadened and the bleach amplitude is significantly reduced.

Fluence dependent measurements (from $<N_X> = 0.2$ to 2.1) are shown for the neutral and charged NPL solid in Figure 4(g) and (h) respectively. Narrow bleach features are observed for all excitation densities in the neutral film. There is no significant broadening or spectral shift of the absorption features, but the HH transition is fully bleached at the highest fluence. In contrast, in the charged film, we see significant broadening of all transient absorption features, a strong redshift of the HH transition (*i.e.*, stimulated emission at the HH energy) and a strong reduction of all bleach amplitudes. The *ΔA* signal at ~2.0 eV is presumably stimulated emission/state filling of free carrier transitions, since light holes will relax to the HH band, and consequently no SE of the LH exciton is expected. This bleach feature shows a significant blueshift at the highest excitation density.

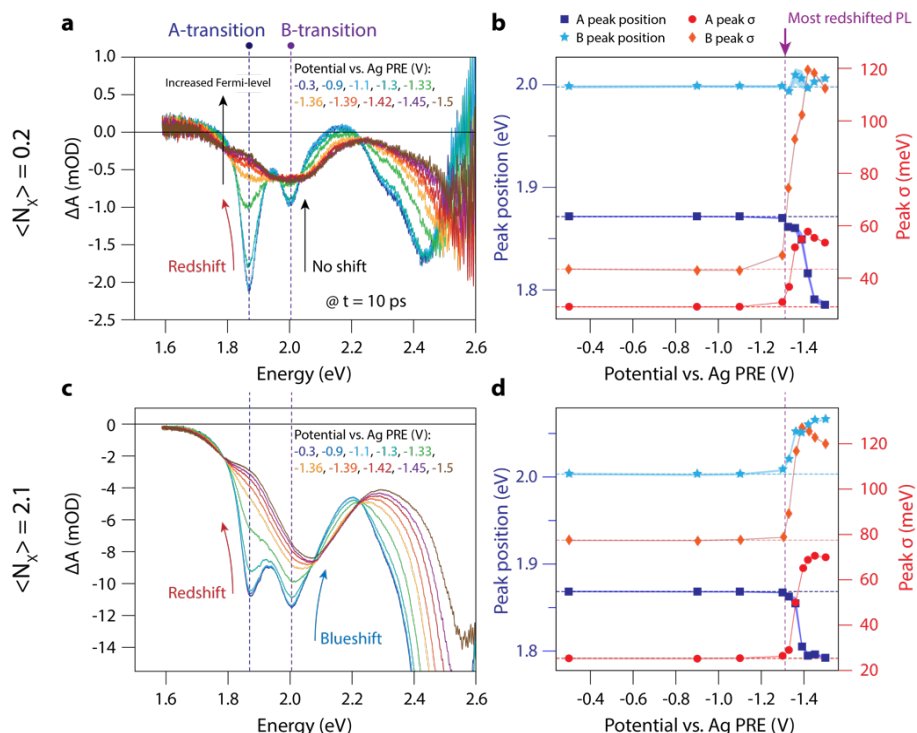

**Figure 5: TA spectra as a function of potential and pump-probe delay time for low and high excitation density (<$N_X$> = 0.2 [a,b] and 2.1 [c,d]). (a)** TA spectra at 10 ps pump-probe delay for various applied electrochemical potentials at low excitation density (<$N_X$> = 0.2). Notice how the higher energy features (related to the CdS shell) are blueshifted gradually and quenched as a function of applied potential **(b)** Fitted peak position and standard deviation of the A and B transitions as a function of applied electrochemical potential. **(c)** TA spectra at 10 ps pump-probe delay for various applied electrochemical potentials at high excitation density (<$N_X$> = 2.1). **(d)** Fitted peak position and standard deviation of the A and B transitions as a function of applied electrochemical potential.

Figure 5(a) presents ΔA spectra for varying applied electrochemical potentials at a pump-probe delay time of 10 ps and for low (<$N_X$> = 0.2) excitation density. We have fitted the bleach features at the HH and LH energy (which we label 'A' and 'B', respectively) with the sum of two Gaussians and plot the resulting peak position and peak width (standard deviation) as a function of potential in Figure 5(b). The A feature redshifts by 90 meV upon decreasing the potential and broadens from 29 meV to 58 meV. Screening of the binding energy would result in a blueshift of the HH transition, so the observed redshift must come from bandgap renormalization and/or the formation of charged excitonic species with an increased attractive interaction compared to the neutral exciton. In contrast, the ΔA peak at ~2.0 eV (feature B), which we assign to overlapping LH exciton and free carrier transitions, does not show an appreciable shift in the peak position, but it broadens from 42 meV to 119 meV.

At higher excitation density ($<N_X>$ = 2.1), shown in Figure 5(c,d), we observe similar trends; the A transition redshifts by 70 meV and broadens from 25 to 73 meV. However, in contrast to low-excitation densities, the peak around 2.0 eV does not stay at a constant energy, but blueshifts by 45 meV (next to broadening from 77 meV to 120 meV). The potential where the PL is most red-shifted, from the SEC-PL data in Figure 2, coincides with the starting potential [purple arrow on top of Figure 5(b)] of the shift of bleach features A and B presented in Figure 5(b,d). The position of the PL energy does not coincide with the free-carrier energy, so we suspect that the leftover PL at negative potentials comes predominantly from a fraction of NPLs in the film that is not charged. This could explain the smaller shifts (<10 meV) of the PL energy compared to the shift of the TA signals (70 – 100 meV).

The ultrafast spectro-electrochemistry results shown in Figures 4 and 5 show that electron injection leads to strong broadening of the absorption transitions, similar to the broadening observed in the PL upon charging (Figure 3a and 3e). What we certainly do not observe is the formation of sharp exciton-like bleach features in the charged NPL film, as is predicted by the scenario shown in Figure 4(a). Note that SE is observed at the HH energy, but that it is much broader than the HH exciton absorption in the neutral film. This confirms that severe broadening of the transitions takes place in the electron-charged NPLs in addition to state filling.

We consider that there are several causes for this broadening. First, scattering with injected electrons could cause rapid dephasing of the excitons which results in lifetime broadening of the excitonic transitions. In addition, exchange interactions between the electrochemically injected electrons and the excitations by the pump and probe pulses could induce broadening. A simple way to look at this effect is to consider that rather than forming a neutral excited species, in the charged NPL there is a broad distribution of charged excited carriers, with a resulting broad spectrum.

ZERO THRESHOLD OPTICAL GAIN IN DOPED NPL SOLIDS

We previously demonstrated that the light-amplifying properties in QD solids can be quantitatively controlled via electrochemical $n$ doping.[7] Since NPLs have larger material gain coefficients than QDs[11] and show very low ASE thresholds[8,20,31–33], we expect a superior performance of a an $n$ doped film of NPLs compared to QDs for lasing applications.

Figure 6(a) presents gain maps of neutral (top row) and charged (bottom row, at -1.5V vs. PRE) NPL films. Upon increasing the excitation density, the neutral film starts to exhibit optical gain around $<N_X> = 0.9$ excitons per platelet, with a gain lifetime of ~600 ps. The development of gain in the film of NPLs is very similar to that of NPLs in solution shown in Fig. 2 above, albeit with narrower HH and free-carrier gain bands. Upon doping the film with electrons, shown in the bottom row of Figure 6(a), we observe optical gain already at the lowest excitation density of $<N_X> = 0.2$ excitons per NPL. When we further increase the fluence, the gain increases strongly. The gain features do not develop in well-defined, separated gain bands from distinct transitions, but appear as broad features in the gain map.

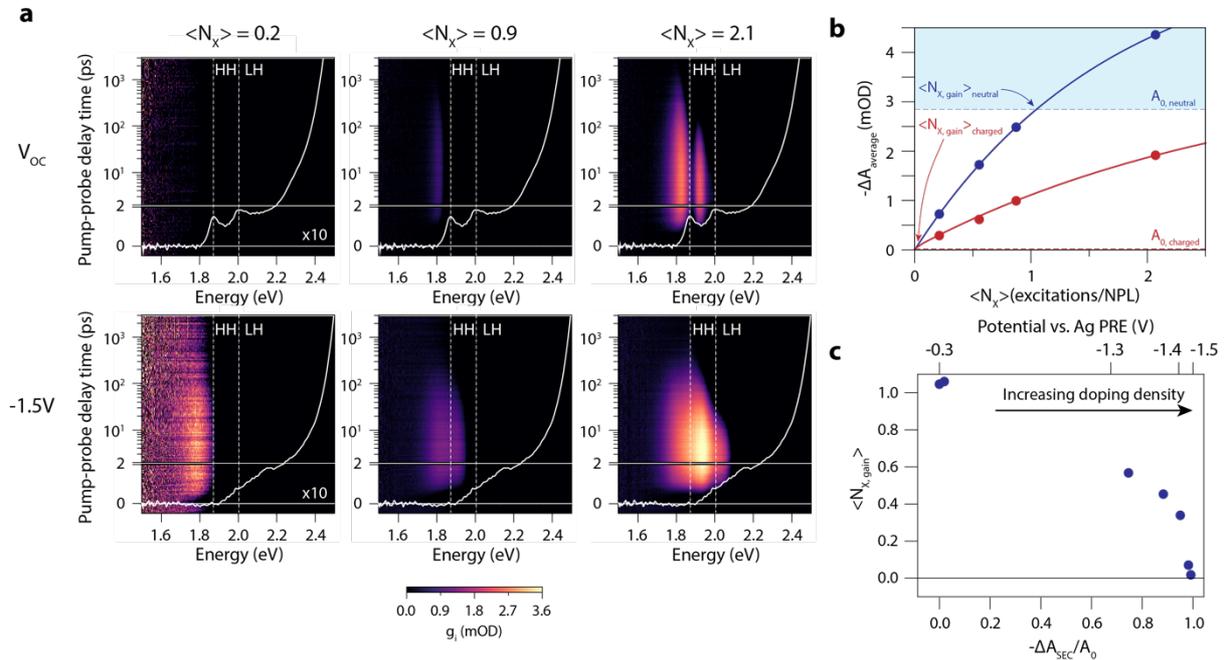

**Figure 6: Optical gain in neutral and charged NPL solids. (a)** Gain maps for different $<N_X>$ and doping densities (neutral: top row; *n* charged: bottom row). The maximum gain is achieved in between the HH and LH transition in the charged NPL film. **(b)** Spectrally averaged transient absorption vs. excitation density. Interpolation or extrapolation is used to determine the spectrally averaged gain threshold for the undoped and doped NPL film. **(c)** The optical gain threshold as a function of electrochemical fractional bleach. The transient absorption is spectrally integrated over the HH transition to account for spectral shifts. The gain threshold is reduced from $<N_{X,\,gain}>$ = 1.05 to 0.02 excitons per NPL upon *n* charging the NPL film.

As we did above for NPLs in solution (Fig 2e), we determine the spectrally averaged gain threshold near the band-edge transitions (averaged between 1.7 and 1.9 eV) for the neutral and charged NPL films, see Figure 6(b). The gain threshold is obtained by interpolating or extrapolating the curve in Figure 6(b) to determine the excitation density where $-\Delta A_{average} = A_{0,average}$. We do this for decreasing potentials and plot the resulting gain threshold *vs.* the fractional electrochemical absorption bleach $\Delta A_{SEC}/A_0$ in Figure 6(c). Upon charging the film, the spectrally integrated gain threshold is reduced from $\langle N_{X, gain}\rangle = 1.05$ excitons per NPL at $V_{OC}$, down to $\langle N_{X, gain}\rangle = 0.02$ excitons per NPL (corresponding to a fluence of 170 nJ/cm²/pulse) at -1.5V, demonstrating the near-complete removal of the threshold for light amplification. The spectrally integrated optical gain lifetime decreases from 1 ns in the neutral film to 500 ps in the charged NPL film.

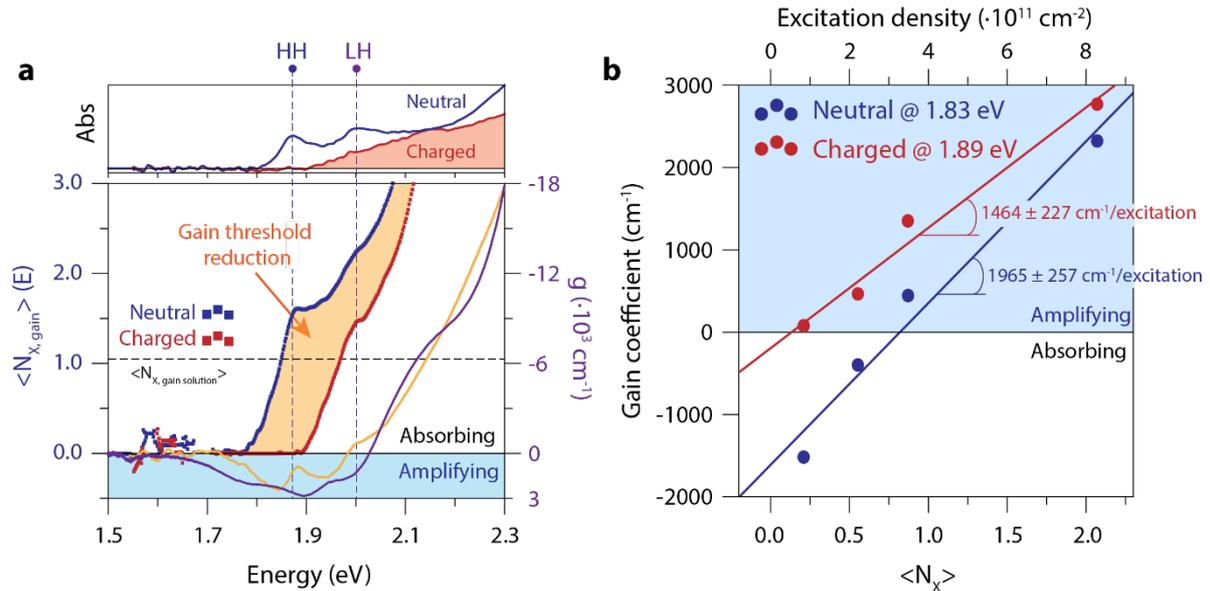

**Figure 7: Quantifying optical gain in electrochemically doped NPL solids. (a)** Gain threshold spectra for the neutral (blue) and doped (red) NPL solid. The yellow area indicates the region in which the gain threshold is significantly reduced by electrochemical doping. The top panel shows the steady-state absorption spectra of the film in the neutral and charged state. For comparison, also the gain-coefficient spectra are shown (yellow line = neutral, purple line = charged). **(b)** Gain coefficients as a function of excitation density at the energies with the maximum gain coefficient (1.83 eV for the neutral NPL film, 1.89 eV for the charged NPL film). To calculate the gain coefficients, we measured the film thickness to be 50.8 ± 15.5 nm and assumed a volume fraction of NPLs of 0.5 in the film.

So far, we focused on the spectrally averaged gain-threshold, since this parameter is the most relevant for discussing the photophysics. For most practical applications, the gain threshold at a single wavelength is more relevant. We plot the gain threshold spectrum in Figure 7(a) for neutral (blue squares) and charged (red squares) films of NPLs (*i.e.*, the gain threshold vs. energy, equivalent to the analysis of the data Figure 2(f)). For comparison, we also show gain coefficient spectra (yellow = neutral, purple = charged). These are obtained from the excited state absorption spectra and the film thickness d = 50.8 ± 15.5 nm (see Methods), as $g = \frac{-A^*\ln(10)}{d}$. Charging the NPLs strongly reduced the gain threshold over a wide energy range (yellow shaded area). The threshold vanishes between 1.8 and 1.9 eV, a spectral region where there is significant light amplification, as shown by the gain spectrum of the charged film (purple solid line).

In Figure 7(b) we plot the gain coefficients at the energies where the neutral and charged NPL film have the maximum amount of gain, 1.83 eV and 1.89 eV respectively, as a function of excitation density. The gain coefficients of the charged NPLs are higher than the neutral NPLs, and reach values up to 2800 cm$^{-1}$ with no sign of saturation. The material gain per excitation, i.e., the slope of the gain coefficient vs $<N_X>$ curve at these energies, are 1965 ± 256 cm$^{-1}$/excitation for the neutral NPL film and 1464 ± 227 cm$^{-1}$/excitation for the charged NPL film. A balance between gain threshold, gain coefficient and gain lifetime has to be sought in order to find the ideal doping density for device purposes. Compared to QDs[7], the NPLs have three to four times higher gain coefficients, and are able to reach lower gain thresholds. Furthermore, the gain coefficients are higher than commonly used erbium-doped fiber amplifiers (10$^{-2}$–10$^{-3}$ cm$^{-1}$) and on par with those of epitaxially grown III-V semiconductors (10$^3$ cm$^{-1}$)[34,35]. These results demonstrate the promising performance of charged NPL films as light amplifying material and highlight the possibilities of using electrochemically doped NPL solids as gain medium in low-threshold lasing devices.

CONCLUSIONS

To summarize, we demonstrate the complex interplay of state-filling and screening of Coulomb and exchange interactions in CdSe/CdS/ZnS NPL solids, which results in zero-threshold optical gain. Both photoexcitation and electrochemical *n* doping of NPLs, leads to a complete bleach of the excitonic transitions. We show that photoexcitation in both neutral and charged core-shell-shell NPLs leads to the formation of free-carriers and that excitonic effects in these systems are weak.

Furthermore, we demonstrated that we can controllably reduce the spectrally averaged threshold for optical gain upon charging the NPL solid from 1 excitation per NPL ($4 \cdot 10^{11}$ cm$^{-2}$) down to 0.02 excitations per NPL ($8 \cdot 10^{9}$ cm$^{-2}$). The NPL light amplifying properties are superior to QDs, as the optical gain thresholds are lower and the gain coefficients are up to 4 times higher. Finally, we model the optical gain in our NPLs, and show that there are optimal lateral sizes of the NPLs that should lead to a minimum threshold density for optical gain. Our results show that the underlying physical effects that govern gain in doped NPLs are a complicated mixture of state filling and screening effects. In short, even though their photophysics is complex, *n* doped NPLs show extremely efficient optical gain which makes them highly promising materials for light amplification and lasing.

**Supporting Information available.** Description of the synthesis of the NPLs employed in this work. Description of the techniques used to characterize the NPLs (TEM, steady-state absorption, PL). Description of the setups used for the TA and spectroelectrochemical measurements. SEC measurements on core and core-shell NPLs, derivation of the Heisenberg model, determination of the absorption cross section using various methods. This material is available free of charge via the Internet at http://pubs.acs.org.

**Competing financial interests.** The authors declare no competing financial interests.

**Corresponding authors.**   Arjan J. Houtepen: a.j.houtepen@tudelft.nl

Jaco J. Geuchies: jaco.geuchies@gmail.com

**Acknowledgements.** AJH and JJG gratefully acknowledge financial support from the European Research Council Horizon 2020 ERC Grant Agreement No. 678004 (Doping on Demand). GG acknowledges financial support from STW (Project No. 13903, Stable and Non-Toxic Nanocrystal Solar Cells). The authors gratefully acknowledge fruitful discussion with Prof. Pieter Geiregat from Gent University.

REFERENCES

(1) Brechbühler, R.; Vonk, S. J. W.; Aellen, M.; Lassaline, N.; Keitel, R. C.; Cocina, A.; Rossinelli, A. A.; Rabouw, F. T.; Norris, D. J. Compact Plasmonic Distributed-Feedback Lasers as Dark Sources of Surface Plasmon Polaritons. *ACS Nano* **2021**, *15* (6), 9935–9944. https://doi.org/10.1021/acsnano.1c01338.

(2) Rossinelli, A. A.; Rojo, H.; Mule, A. S.; Aellen, M.; Cocina, A.; De Leo, E.; Schäublin, R.; Norris, D. J. Compositional Grading for Efficient and Narrowband Emission in CdSe-Based Core/Shell Nanoplatelets. *Chem. Mater.* **2019**, *31* (22), 9567–9578. https://doi.org/10.1021/acs.chemmater.9b04220.

(3) Klimov, V. I. Spectral and Dynamical Properties of Multiexcitons in Semiconductor Nanocrystals. *Annu. Rev. Phys. Chem.* **2007**, *58* (1), 635–673. https://doi.org/10.1146/annurev.physchem.58.032806.104537.

(4) Wang, C.; Wehrenberg, B. L.; Woo, C. Y.; Guyot-Sionnest, P. Light Emission and Amplification in Charged CdSe Quantum Dots. *J. Phys. Chem. B* **2004**, *108* (26), 9027–9031. https://doi.org/10.1021/jp0489830.

(5) Wu, K.; Park, Y.-S.; Lim, J.; Klimov, V. I. Towards Zero-Threshold Optical Gain Using Charged Semiconductor Quantum Dots. *Nat. Nanotechnol.* **2017**, *12* (12), 1140–1147. https://doi.org/10.1038/nnano.2017.189.

(6) Kozlov, O. V.; Park, Y. S.; Roh, J.; Fedin, I.; Nakotte, T.; Klimov, V. I. Sub–Single-Exciton Lasing Using Charged Quantum Dots Coupled to a Distributed Feedback Cavity. *Science (80-. ).* **2019**, *365* (6454), 672–675.

https://doi.org/10.1126/science.aax3489.

(7) Geuchies, J. J.; Brynjarsson, B.; Grimaldi, G.; Gudjonsdottir, S.; Van Der Stam, W.; Evers, W. H.; Houtepen, A. J. Quantitative Electrochemical Control over Optical Gain in Quantum-Dot Solids. *ACS Nano* **2021**, *15* (1), 377–386. https://doi.org/10.1021/acsnano.0c07365.

(8) Guzelturk, B.; Kelestemur, Y.; Olutas, M.; Delikanli, S.; Demir, H. V. Amplified Spontaneous Emission and Lasing in Colloidal Nanoplatelets. *ACS Nano* **2014**, *8* (7), 6599–6605. https://doi.org/10.1021/nn5022296.

(9) Guzelturk, B.; Pelton, M.; Olutas, M.; Demir, H. V. Giant Modal Gain Coefficients in Colloidal II–VI Nanoplatelets. *Nano Lett.* **2019**, *19* (1), 277–282. https://doi.org/10.1021/acs.nanolett.8b03891.

(10) Geiregat, P.; Tomar, R.; Chen, K.; Singh, S.; Hodgkiss, J. M.; Hens, Z. Thermodynamic Equilibrium between Excitons and Excitonic Molecules Dictates Optical Gain in Colloidal CdSe Quantum Wells. *J. Phys. Chem. Lett.* **2019**, *10* (13), 3637–3644. https://doi.org/10.1021/acs.jpclett.9b01607.

(11) Tomar, R.; Kulkarni, A.; Chen, K.; Singh, S.; Van Thourhout, D.; Hodgkiss, J. M.; Siebbeles, L. D. A.; Hens, Z.; Geiregat, P. Charge Carrier Cooling Bottleneck Opens Up Nonexcitonic Gain Mechanisms in Colloidal CdSe Quantum Wells. *J. Phys. Chem. C* **2019**, *123* (14), 9640–9650. https://doi.org/10.1021/acs.jpcc.9b02085.

(12) Schmitt-Rink, S.; Chemla, D. S.; Miller, D. A. B. Theory of Transient Excitonic Optical Nonlinearities in Semiconductor Quantum-Well Structures. *Phys. Rev. B* **1985**,

*32* (10), 6601–6609. https://doi.org/10.1103/PhysRevB.32.6601.

(13) Naeem, A.; Masia, F.; Christodoulou, S.; Moreels, I.; Borri, P.; Langbein, W. Giant Exciton Oscillator Strength and Radiatively Limited Dephasing in Two-Dimensional Platelets. *Phys. Rev. B* **2015**, *91* (12), 121302. https://doi.org/10.1103/PhysRevB.91.121302.

(14) Geiregat, P.; Rodá, C.; Tanghe, I.; Singh, S.; Di Giacomo, A.; Lebrun, D.; Grimaldi, G.; Maes, J.; Van Thourhout, D.; Moreels, I.; Houtepen, A. J.; Hens, Z. Localization-Limited Exciton Oscillator Strength in Colloidal CdSe Nanoplatelets Revealed by the Optically Induced Stark Effect. *Light Sci. Appl.* **2021**, *10* (1), 112. https://doi.org/10.1038/s41377-021-00548-z.

(15) Chernikov, A.; Ruppert, C.; Hill, H. M.; Rigosi, A. F.; Heinz, T. F. Population Inversion and Giant Bandgap Renormalization in Atomically Thin WS 2 Layers. *Nat. Photonics* **2015**, *9* (7), 466–470. https://doi.org/10.1038/nphoton.2015.104.

(16) Brumberg, A.; Harvey, S. M.; Philbin, J. P.; Diroll, B. T.; Lee, B.; Crooker, S. A.; Wasielewski, M. R.; Rabani, E.; Schaller, R. D. Determination of the In-Plane Exciton Radius in 2D CdSe Nanoplatelets via Magneto-Optical Spectroscopy. *ACS Nano* **2019**, *13* (8), 8589–8596. https://doi.org/10.1021/acsnano.9b02008.

(17) Kelestemur, Y.; Shynkarenko, Y.; Anni, M.; Yakunin, S.; De Giorgi, M. L.; Kovalenko, M. V. Colloidal CdSe Quantum Wells with Graded Shell Composition for Low-Threshold Amplified Spontaneous Emission and Highly Efficient Electroluminescence. *ACS Nano* **2019**, *13* (12), 13899–13909. https://doi.org/10.1021/acsnano.9b05313.

(18) Rossinelli, A. A.; Riedinger, A.; Marqués-Gallego, P.; Knüsel, P. N.; Antolinez, F. V.; Norris, D. J. High-Temperature Growth of Thick-Shell CdSe/CdS Core/Shell Nanoplatelets. *Chem. Commun.* **2017**, *53* (71), 9938–9941. https://doi.org/10.1039/C7CC04503D.

(19) Van Der Stam, W.; Grimaldi, G.; Geuchies, J. J.; Gudjonsdottir, S.; Van Uffelen, P. T.; Van Overeem, M.; Brynjarsson, B.; Kirkwood, N.; Houtepen, A. J. Electrochemical Modulation of the Photophysics of Surface-Localized Trap States in Core/Shell/(Shell) Quantum Dot Films. *Chem. Mater.* **2019**, *31* (20), 8484–8493. https://doi.org/10.1021/acs.chemmater.9b02908.

(20) Grim, J. Q.; Christodoulou, S.; Di Stasio, F.; Krahne, R.; Cingolani, R.; Manna, L.; Moreels, I. Continuous-Wave Biexciton Lasing at Room Temperature Using Solution-Processed Quantum Wells. *Nat. Nanotechnol.* **2014**, *9* (11), 891–895. https://doi.org/10.1038/nnano.2014.213.

(21) Pokutnii, S. I. Exciton Binding Energy in Semiconductor Quantum Dots. *Semiconductors* **2010**, *44* (4), 488–493. https://doi.org/10.1134/S1063782610040147.

(22) Yu, P.; Cardona, M. *Fundamentals of Semiconductors – Physics and Materials Properties*, 3rd ed.; Springer, 2010. https://doi.org/10.1524/zpch.1997.198.part_1_2.275.

(23) Sturge, M. D. Optical Absorption of Gallium Arsenide between 0.6 and 2.75 EV. *Phys. Rev.* **1962**, *127* (3), 768–773. https://doi.org/10.1103/PhysRev.127.768.

(24) Tarucha, S.; Okamoto, H.; Iwasa, Y.; Miura, N. Exciton Binding Energy in GaAs

Quantum Wells Deduced from Magneto-Optical Absorption Measurement. *Solid State Commun.* **1984**, *52* (9), 815–819. https://doi.org/10.1016/0038-1098(84)90012-7.

(25) Van Der Bok, J. C.; Dekker, D. M.; Peerlings, M. L. J.; Salzmann, B. B. V.; Meijerink, A. Luminescence Line Broadening of CdSe Nanoplatelets and Quantum Dots for Application in W-LEDs. *J. Phys. Chem. C* **2020**, *124* (22), 12153–12160. https://doi.org/10.1021/acs.jpcc.0c03048.

(26) Van Der Stam, W.; Du Fossé, I.; Grimaldi, G.; Monchen, J. O. V.; Kirkwood, N.; Houtepen, A. J. Spectroelectrochemical Signatures of Surface Trap Passivation on CdTe Nanocrystals. *Chem. Mater.* **2018**, *30* (21), 8052–8061. https://doi.org/10.1021/acs.chemmater.8b03893.

(27) du Fossé, I.; ten Brinck, S.; Infante, I.; Houtepen, A. J. The Role of Surface Reduction in the Formation of Traps in N-Doped II-VI Semiconductor Nanocrystals: How to Charge without Reducing the Surface. *Chem. Mater.* **2019**, acs.chemmater.9b01395. https://doi.org/10.1021/acs.chemmater.9b01395.

(28) Jha, P. P.; Guyot-Sionnest, P. Photoluminescence Switching of Charged Quantum Dot Films. *J. Phys. Chem. C* **2007**, *111* (42), 15440–15445. https://doi.org/10.1021/jp075042p.

(29) Antolinez, F. V.; Rabouw, F. T.; Rossinelli, A. A.; Cui, J.; Norris, D. J. Observation of Electron Shakeup in CdSe/CdS Core/Shell Nanoplatelets. *Nano Lett.* **2019**, *19* (12), 8495–8502. https://doi.org/10.1021/acs.nanolett.9b02856.

(30) Gudjonsdottir, S.; Koopman, C.; Houtepen, A. J. Enhancing the Stability of the

Electron Density in Electrochemically Doped ZnO Quantum Dots. *J. Chem. Phys.* **2019**, *151* (14), 144708. https://doi.org/10.1063/1.5124534.

(31) Olutas, M.; Guzelturk, B.; Kelestemur, Y.; Yeltik, A.; Delikanli, S.; Demir, H. V. Lateral Size-Dependent Spontaneous and Stimulated Emission Properties in Colloidal CdSe Nanoplatelets. *ACS Nano* **2015**, *9* (5), 5041–5050. https://doi.org/10.1021/acsnano.5b01927.

(32) Li, Q.; Lian, T. A Model for Optical Gain in Colloidal Nanoplatelets. *Chem. Sci.* **2018**, *9* (3), 728–734. https://doi.org/10.1039/c7sc04294a.

(33) She, C.; Fedin, I.; Dolzhnikov, D. S.; Dahlberg, P. D.; Engel, G. S.; Schaller, R. D.; Talapin, D. V. Red, Yellow, Green, and Blue Amplified Spontaneous Emission and Lasing Using Colloidal CdSe Nanoplatelets. *ACS Nano* **2015**, *9* (10), 9475–9485. https://doi.org/10.1021/acsnano.5b02509.

(34) Kittel, C. *Introduction to Solid State Physics*; Wiley, 2005.

(35) Jiang, S.; Luo, T.; Hwang, B.-C.; Smekatala, F.; Seneschal, K.; Lucas, J.; Peyghambarian, N. Er3+-Doped Phosphate Glasses for Fiber Amplifiers with High Gain per Unit Length. *J. Non. Cryst. Solids* **2000**, *263–264*, 364–368. https://doi.org/10.1016/S0022-3093(99)00646-8.

# Supplementary Information
## for
Zero-threshold optical gain in electrochemically doped nanoplatelets and the physics behind it

Jaco J. Geuchies[†*], Robbert Dijkhuizen[†], Marijn Koel[†], Gianluca Grimaldi[†$], Indy du Fossé[†], Wiel H. Evers[†], Zeger Hens[#], Arjan J. Houtepen[†]

† Optoelectronic Materials Section, Faculty of Applied Sciences, Delft University of Technology, Van der Maasweg 9, 2629 HAZ Delft, The Netherlands

[#]Department of Chemistry and Center for Nano and Biophotonics, Ghent University, 9000 Ghent, Belgium;

* Current address: Max Planck Institute for Polymer Research, 55128 Mainz, Germany.

$ Current address: Center for Nanophotonics, AMOLF, 1098 XG Amsterdam, the Netherlands / Cavendish Laboratory, University of Cambridge, CB2 1TN Cambridge, United Kingdom.

# Table of Contents

## Section S1 - Methods

**Materials:** 1-Octadecene (ODE) [90%, anhydrous, degassed, Sigma Aldrich], Sodium Myristate (Na(myr)) [Sigma Aldrich], Cadmium Nitrate Tetrahydrate (Cd(NO$_3$)$_2$·4H$_2$O)[Sigma Aldrich]. Selenium powder (Se) [200 mesh, 99.999%, Sigma Aldrich], Cadmium Acetate (Cd(Ac)$_2$) [99.999%, anhydrous, Chempur], Hexane [96%, anhydrous, TCI], Cadmium Oleate (Cd(Oleate)$_2$) [0.076M in ODE, preparation described bbelow], Methanol [99.8%, anhydrous, Sigma Aldrich], Buthanol [96.0%, anhydrous Sigma Aldrich], Oleylamine[80-90%, Sigma Aldrich], 1-Octane-Thiol [98.5%, Sigma Aldrich], Acetonitrile [99.8% anhydrous,[Sigma Aldrich], Zinc Oleate (Zn(Oleate)$_2$) [0.086M in ODE, preparation described below], Elemental Sulphur (S) [99.9995%, Sigma Aldrich], Cadmium Acetate Dihydrate (Cd(Ac)$_2$·2H$_2$O) [Fluorochem], Oleic Acid [90% technical grade, degassed, Sigma Aldrich], 1,8-diaminooctane [98%, Sigma Aldrich], 1,8-octanedithiol [97%, Sigma Aldrich), ITO substrate [10x23x0.7mm, SiO$_2$ passivated, Prezisions glas and optik], Cadmiumdichloride (CdCl$_2$) [Sigma Aldrich], Toluene [99.8%, anhydrous, Sigma Aldrich], Ferrocene [98%, Sigma Aldrich], Lithium Perchlorate[99.9%, Sigma Aldrich]. Acetonitril was dried before use in an Innovative Technology PureSolv Micro column. All other chemicals were used as received, unless specifically mentioned.

### Synthesis of bare CdSe nanoplatelets (NPLs).

For the synthesis of the bare CdSe NPLs, we first prepared Cd-myristate and Se-in-ODE precursors as sources for the NPL growth.

**Cadmium precursor: Cd-myristate.** Cd(myr)$_2$ is synthesized by a precipitation reaction of Na(myr) and Cd(NO$_3$)$_2$·4H$_2$O. 1.23g (4mmol) of Cd(NO$_3$)$_2$·4H$_2$O was dissolved in 40ml of Methanol in a beaker. Separately, 3.13g (12.5mmol) of Na(myr) was dissolved in 250ml of Methanol in a beaker. Both beakers where continuously stirred and slightly heated (40°C) to speed up the dissolving process. The time required to fully dissolve the Na(myr) and $Cd(NO_3)_2 \cdot 4H_2O$ was usually an hour. To prevent extensive MeOH evaporation, we placed a petridish on top of the beakers with the solutions. After preparation, the dissolved $Cd(NO_3)_2 \cdot 4H_2O$ was gently added to Na(myr) while stirring continuously. The $Cd(myr)_2$ precipitates as a fine, white powder. The entire mixture was filtrated through a Büchner funnel under vacuum. The powder was rinsed twice in the Büchner funnel with 2 times 10 mL of MeOH. The $Cd(myr)_2$ powder was then removed from the funnel and left to dry under vacuum overnight. Afterwards, the $Cd(myr)_2$ was heated to 60°C on a hot plate for an hour inside a glovebox until all the leftover methanol was evaporated and was stored inside the box for later use.

**Selenium precursor: Se-in-ODE.** The Selenium precursor was made by mixing 180mg (2.28mmol) Selenium powder with 15ml of 1-Octadecene (ODE). It should be noted that Selenium powder does not dissolve in ODE, and will therefore remain as a black suspension as room temperature. The suspension should be sonicated for 5 minutes before usage.

**Cd-oleate and Zn-oleate preparation.** 1.32 g of Cd-(acetate)$_2$ was dissolved in 52.4g of ODE and 7.4 g of oleic acid. The mixture was placed under vacuum ($10^{-1}$ mbar) at room temperature for 20 minutes, and heated under vacuum to 120 degrees for three hours. The Cd-oleate solution was transferred air-free into a $N_2$ flushed vial, and placed inside a $N_2$ purged glovebox for further use. Note: the Cd-oleate solution solidifies over time, which can be seen by the formation of a white sluggish gel inside the solution. Before use, the solution is heated to 50°C, until a transparent solution is formed. Usually, the formation of this sluggish gel takes a few days at room temperature. Furthermore, our Cd-oleate solutions usually had a slightly yellow tint.

The Zn-oleate was made in a similar fashion. Zn(II)-(acetate)$_2$ was mixed with 1g of OA, 1.6 mL ODE and 1.6 mL of OLAM. The oleylamine serves as a stabilizing ligand for the Zn-oleate, since this has the tendency to rapidly solidify out of solution at room temperature otherwise. The mixture was heated up in a 20 mL vial inside a nitrogen purged glovebox to 130°C and stored there for further use. Note that the Zn-oleate solution is extremely viscous and should be handled with care when placed into a syringe.

**Synthesis of the CdSe nanoplatelets.** 14ml of ODE, 170mg of Cd(myr)$_2$ and 1ml of Selenium precursor and a magnetic stir bean were added to a 100mL three-neck flask. The mixture was then degassed under vacuum (10$^{-1}$ mbar) at 50°C while stirring continuously for an hour. After degassing, the mixture was heated to 240°C. At 100°C, the mixture turned transparent. At 180°C, the mixture started turning yellow and at 200°C the mixture turned orange. 50mg Cd(Ac)$_2$ was added when the mixture became orange, by quickly removing the nitrogen inlet and poring the Cd-acetate powder using a weighing boat, after which the N$_2$ inlet was placed back. After that, the particles were allowed to grow for 8 minutes and reaction mixture was then cooled to 50°C. Lastly 12 mL of Hexane and 1 mL of 0.076M Cd(Oleate)$_2$ were added. The liquid NPL mixture was then taken out of the flask and placed in a storage vial without exposing it to air.

**Purification of the CdSe nanoplatelets.** The purification of the NPLs was done in two steps. First, the unreacted reagents and quantum dots were separated from the NPL solution. Second, the smaller sized nanoplatelets (460 to 480 nm luminescent NPLs) were separated from the main product (510nm NPLs). For the first purification step, a mixture of 2:1 MeOH:BuOH was added to the NPL solution. This MeOH:BuOH mixture was added drop-wise until the solution turned turbid. The solution was then centrifuged at 3000 rpm for 15 minutes. Afterwards, the precipitate was dissolved in hexane and the supernatant discarded. This process was repeated on the precipitate for three times. For the second purification step, after the last centrifugation step, the precipitate was dissolved in 5 mL hexane and centrifuged at 4000 rpm for 30 minutes. Afterwards, the supernatant was stored in a vial and an absorption spectrum was taken, to ensure the desired product of 510 nm CdSe NPLs was obtained. The two purification steps were then repeated if impurities were still visible in the absorption spectrum.

## **Shell growth of bare CdSe/CdS/ZnS core-shell-shell nanoplatelets.**

The CdSe/CdS/ZnS core-shell-shell synthesis used here was similar to a method by Rossinelli et al.[1] and from Kelestemur et al.[2], where one or more precursors are continuously injected through a syringe pump. The synthesis can be split in two parts:

1. Core-shell-shell synthesis from the core NPLs.

2. Purification of the core-shell-shell NPLs.

**Core-shell-shell synthesis from the core nanoplatelets.** Based on the synthesis here, we obtain 6 monolayers of CdS and 2 monolayers of ZnS on each side of the CdSe NPLs (confirmed by checking the absorption peak positions after CdS shell growth to reported values in literature from Ithurria et al.[3]). To start, 2 mL CdSe core NPLs stock solution, 1.3mL Cd-oleate (0.076M) and 2.5mL ODE were added to a 50mL three-neck flask. The mixture was degassed for 40 minutes at room temperature and then degassed for 20 minutes at 80°C. After degassing, 1mL of oleylamine was added and the temperature was increased to 300°C. Starting at 180°C a solution of 28 μL (0.16mmol) 1-octanethiol dissolved in 4mL ODE was injected at a rate of 1.5mL/h by a syringe pump. After two hours, a solution of 0.04 mmol Zn-oleate in 2 mL ODE was injected at 1.5ml/h, simultaneous with the octanethiol. After completion of the addition of the precursors, the solution was kept at 300°C for 30 more minutes, and afterwards cooled down to room temperature by removing the heating and cooling with an airgun. Lastly, 2.5mL hexane was added to the synthesis product. The synthesized CdSe/CdS/ZnS core-shell-shell NPL solution was taken out of the flask and stored in a nitrogen purged vial.

**Core-shell-shell nanoplatelet purification.** The purification was done in three steps. In the first step, the NPL solution was centrifuged at 4000 rpm for 30 minutes. In the second step, the precipitate was re-dispersed in 5mL hexane and a few droplets of oleylamine. The re-dispersed NPL solution was then again centrifuged at 4000 rpm for 30 minutes. In the third step a 2:1 MeOH:BuOH mixture was added to the supernatant of the mixture and was then once again centrifuged at 4000 rpm for 30 minutes. The last two steps were repeated twice for further purification. Afterwards, the precipitate was re-dispersed in 5 mL hexane and stored in a vial inside a nitrogen purged glovebox. Absorption and PL spectra were recorded and further purification was done by repeating the last step if necessary, based on these spectra.

**QDs-on-ITO film preparation by dipcoating.** We used the core-shell-shell NPL solution as obtained by the procedure above. Before dipcoating, the ITO slides were cleaned by sonication in isopropanol and rinsing with ethanol and acetone, followed by drying with an airgun. The slides were placed inside a UV-Ozone cleaner for 30 minutes prior to dipcoating, to increase the wetting of the NPL solution on the ITO. The dipcoating was performed using a dipcoater from Nima Technology, while our glovebox was set on 'purge mode' to ensure that the solvents are extracted from the box efficiently by the continuous flow of $N_2$ gas. The ITO slides were consecutively dipped for 30 seconds in a solution of the colloidal NPLs (1), a solution of 0.1M 1,8-octanedithiol in MeOH (2) and pure MeOH (3) to remove an excess of unbound ligands. In between the consecutive dipping steps, the solvent (hexane, MeOH) was allowed to evaporate for 30 seconds. These steps were repeated at least 20 times to ensure the build-up of a decently thick film. Roughly $1/3^{rd}$ of the entire substrate was left uncoated (and cleaned with a cotton swab and ethanol later) to ensure good contact with the electrodes in our electrochemical experiments (see below).

**Steady state absorption and photoluminescence measurements.** Absorption spectra were measured on a double-beam PerkinElmer Lambda 1050 UV/Vis spectrometer; in case of the NPL films on ITO, the sample was measured inside an integrating sphere and an empty ITO was measured separately for background correction. Photoluminescence spectra were recorded on an Edinburgh Instruments FLS980 spectrofluorimeter equipped with double grating monochromators for both excitation and emission paths and a 450 W Xenon lamp as an excitation source.

**Transmission Electron Microscopy (TEM).** TEM images were acquired using a JEOL JEM-1400 plus TEM microscope operating at 120 kV. Samples for TEM imaging were prepared by dropcasting a dilute solution of NPLs onto a Formvar and carbon coated copper (400-mesh) TEM grid.

**fs-Transient Absorption (TA) spectroscopy.** fs-TA measurements are performed on solutions of the CdSe(/CdS/ZnS) NPLs in hexane or toluene, loaded inside an air-tight cuvet inside a nitrogen purged glovebox. A Yb-KGW oscillator (Light Conversion, Pharos SP) is used to produce 180 fs photon pulses with a wavelength of 1028 nm and at a frequency of 5 kHz. The pump beam is obtained by sending the fundamental beam through an Optical Parametric Amplifier (OPA) equipped with a second harmonic module (Light Conversion, Orpheus), performing non-linear frequency mixing and producing an output beam whose wavelength can be tuned in the 310-1330 nm window. A small fraction of the fundamental beam power is used to produce a broadband probe spectrum (480-1600 nm), by supercontinuum generation in a sapphire crystal. The pump beam is transmitted through a mechanical chopper operating at 2.5 kHz, allowing one in every two pump pulses to be transmitted. Pump and probe beam overlap at the sample position with a small angle (roughly 8°), and with a relative time delay controlled by an automated delay-stage. After transmission through the sample, the pump beam is dumped while the probe is collected at a detector (Ultrafast Systems, Helios). During the experiments, we make sure the pump and probe beam have orthogonal polarizations (i.e. one of them is vertically polarized, the other horizontally), to reduce the influence of pump scattering into our detector. The differential absorbance is obtained via $\Delta A = log(I_{on}/I_{off})$, where $I$ is the probe light incident on the detector with either pump on or pump off. TA data are corrected for probe-chirp via a polynomial correction to the coherent artifact. Pump photon fluence was estimated by measuring the power with a thermopile sensor (Coherent, PS19Q) and obtaining the beamshape with a beamprofiler.

We also measure transient reflection (TR) spectra to obtain the true change in absorption in transient transmission experiments.

**Photoluminescence quantum yield (PLQY) measurements.** We measured the PLQY of the NPL dispersions with respect to a Rhodamine 101 solution in ethanol. The PLQY was calculated using the following equation;

$$PLQY = PLQY_{Rhodamine\ 101} \frac{I^{PL}_{QD\ solution}}{I^{PL}_{Rhodamine\ 101}} \frac{f_{Rhodamine\ 101}}{f_{QD\ solution}} \left(\frac{n_{hexane}}{n_{ethanol}}\right)^2$$

Where $PLQY_{rhodamine\ 101}$ is set to be 95%, $I^{PL}$ is the intensity of the photoluminescence signal of either the QD solution or the Rhodamine 101 solution, $n_{hexane/ethanol}$ is the refractive index of hexane or ethanol at 530 nm (1.377 and 1.3630) and $f_x$ is the fraction of absorbed light of species x, calculated as $f_x = 1 - 10^{-OD_x}$, where $OD_x$ is the optical density of the solution containing either the QDs or the Rhodamine 101. We determined the PLQY of the CdSe/6CdS/2ZnS core-shell-shell NPLs to be 62%.

**Spectroelectrochemical (SEC) measurements.** The SEC measurements were all performed in a $N_2$ purged glovebox. As an electrolyte, we used an 0.1 M $LiClO_4$ solution in acetonitrile, which was dried with an Innovative Technology PureSolv Micro column. The QD film was immersed in the electrolyte solution, together with a Ag wire pseudoreference electrode and a Pt sheet counter electrode. The potential of the NC film on ITO was controlled with a PGSTAT128N Autolab potentiostat. Changes in the absorption or PL of the NC film as a function of applied potential were recorded simultaneously with a cyclic voltammogram with a fiber-based UV-VIS spectrometer (USB2000, Ocean Optics). For the film, the measurements were started at the open-circuit potential ($V_{OC}$ = -0.3V w.r.t. Ag wire, i.e. -0.75V vs. $Fc/Fc^+$)), while scanning with a rate of 20 mV/s. Unless stated otherwise, all potentials are given w.r.t. the Ag pseudoreference. For SEC measurements combined with fsTA, ultrafast spectroelectrochemistry, we loaded the samples inside a nitrogen purged glovebox into a leak-tight sample holder.

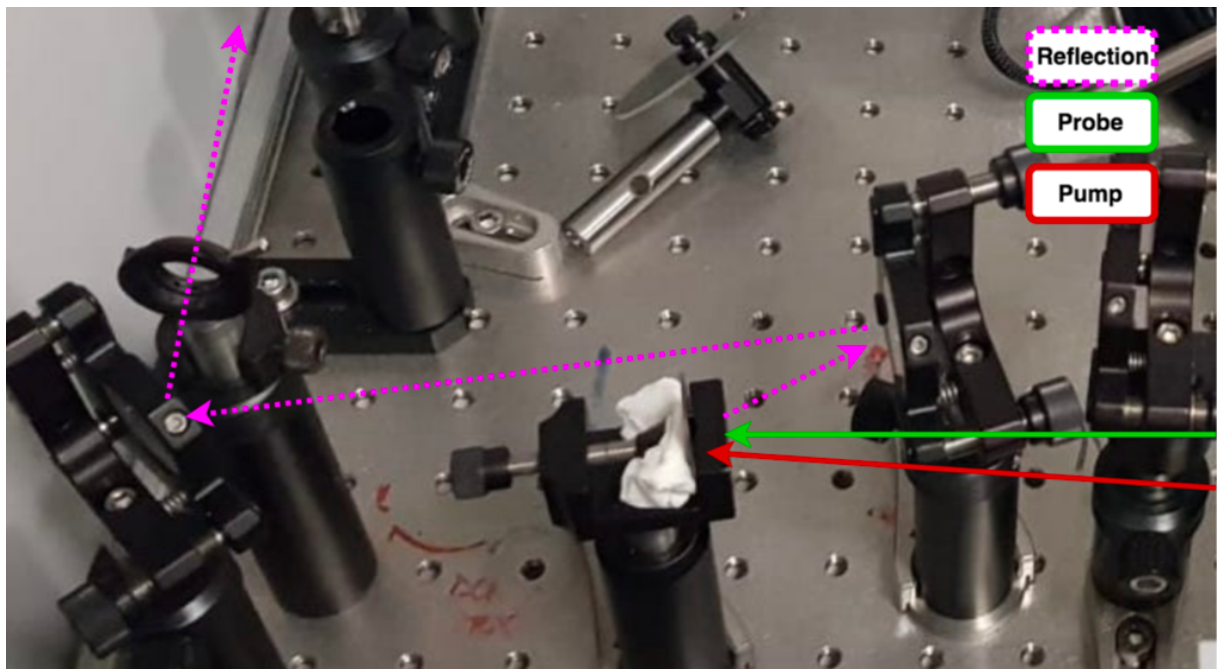

**Figure S1: Alignment of pump and probe for the transient reflection experiments.** The transmitted probe beam is blocked after the sample (not shown in picture). Also here, pump and probe polarizations are orthogonal to each other.

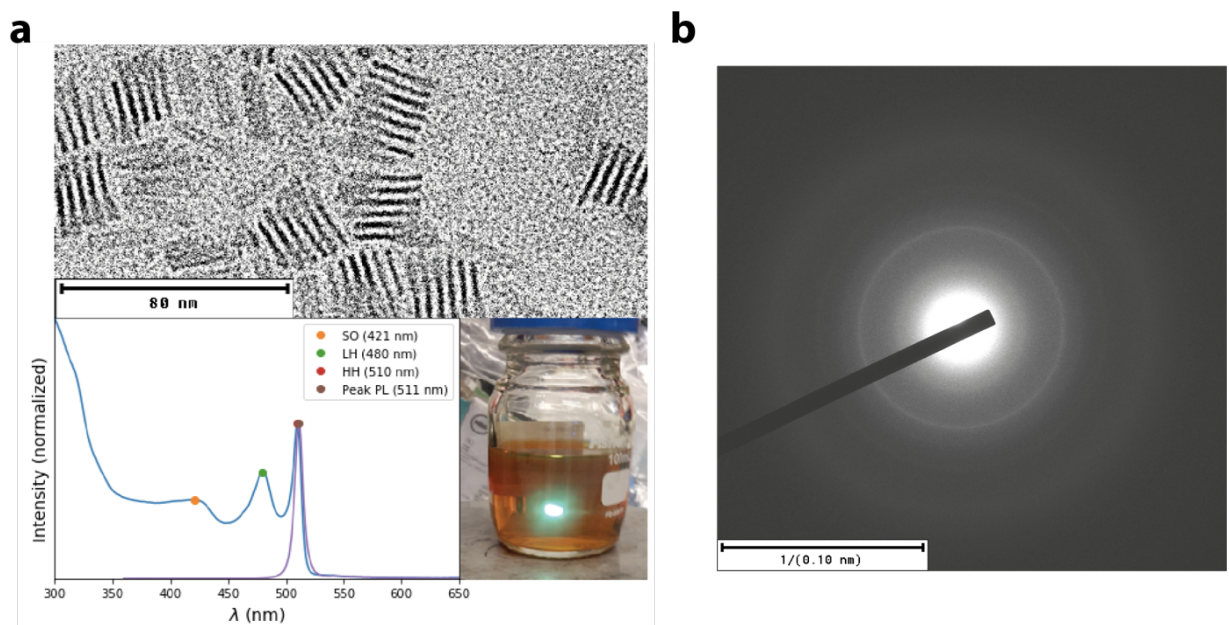

**Figure S2: Optical and structural characterization of the core CdSe NPLs used throughout this study.** (a) Absorption and PL of the core CdSe NPLs, absorbing at 510 nm (bottom). Representative TEM image of the cores (top) and photograph (bottom right) of the NPL stock solution. (b) Electron diffraction pattern of the core CdSe NPLs, indicating a zinc-blende crystal structure (note the strong contrast with a wurtzite crystal structure [prevalent in many quantum-dots], due to the missing two sets of triplet peaks).

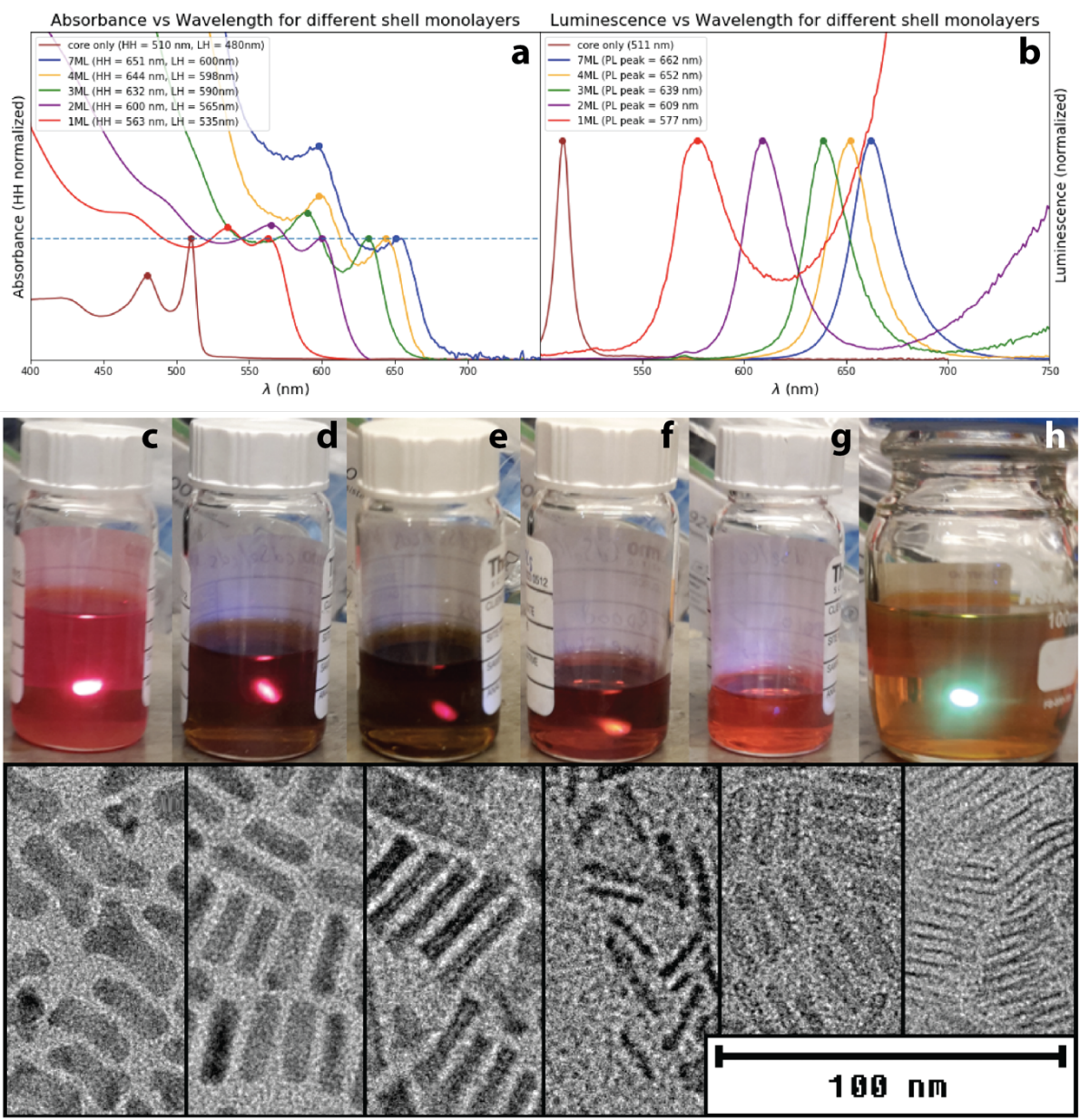

**Figure S3: Optical and structural characterization of CdSe/CdS NPLs grown by continuous injection.** (a) Absorption and (b) PL of the core-shell CdSe/CdS NPLs with different shell thicknesses. The shell thickness was assigned according to work by Ithurria et al.[3]. (c-g) Photographs of the colloidal solutions (top) and TEM images of the CdSe/$n$CdS core-shell NPLs, with $n$ = 7,4,3,2,1 going from left to right respectively, and the core-only CdSe nanoplatelets on the right in (h).

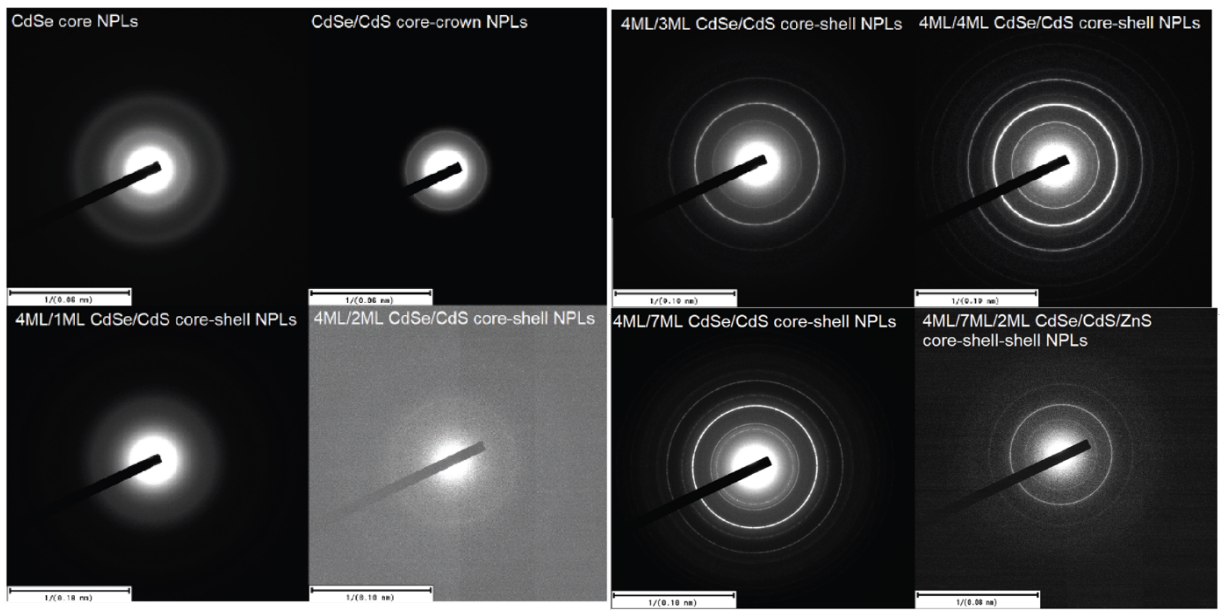

**Figure S4: Electron diffraction on all types of NPLs.** The core-only, core-shell, and core-shell-shell NPLs all keep a zinc-blende crystal structure. Number of monolayers (ML) is indicated in the top of each electron diffractogram.

## Section S2 – Determination of the HH binding energy.

We make a crude estimate of the exciton binding energy vs. the shell thickness by fitting the absorption spectra with an adjusted Elliot-model[4].

$$A(E) = p_X(E) + C(E)$$

Where A(E) is the absorption as a function of energy, $p_X$ the absorption of a quantum well exciton with asymmetric broadening $\eta$ due to localization:

$$p_X(E) = \frac{1}{2\eta}\left[\text{erf}\left(\frac{E-E_0}{\gamma} - \frac{\gamma}{2\eta}\right) + 1\right] \exp\left(\frac{\gamma^2}{4\eta^2} - \frac{E-E_0}{\eta}\right)$$

and C(E) the absorption profile for free-carrier transitions:

$$C(E) = \frac{A_c}{2} \text{erf}\left(\frac{E - E_0 - E_{b,X}}{\gamma_c}\right)$$

Here, $E_0$ and $E_{b,X}$ are the exciton energy and exciton binding energy, respectively. 'Erf' is the error function. The total absorption becomes the sum of both LH and HH contributions:

$$A_{total}(E) = A_{HH} + A_{LH}$$

This model decomposes the absorption spectrum into an excitonic and free-carrier contribution part. We include both the HH and LH exciton and free carrier contributions, and fit the region around the band-edge. For shell thicknesses larger than 3 CdS monolayers, it is very hard to observe the onset of the free-carrier absorption in the absorption spectrum due to the increasing CdS shell absorption. Therefore, we use the 0, 1, 2 and 3 monolayer fits and extrapolate it to the 6 monolayer thickness using a heuristic fit (shown in Fig. 1 of the main text). We note that this is a very rough estimation of the exciton binding energy for the NPLs used throughout this study.

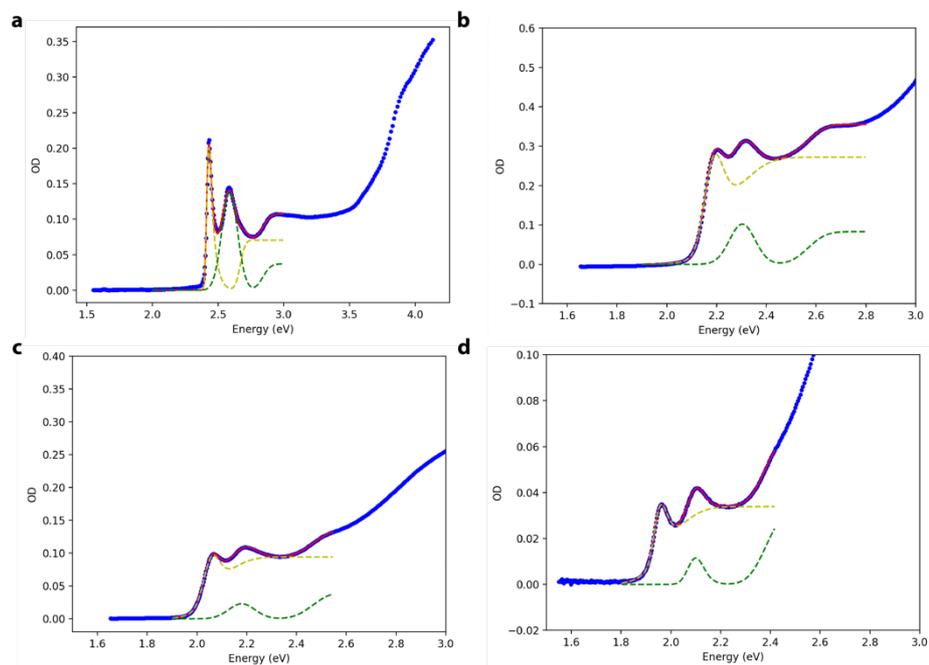

**Figure S5: Fits of the absorption spectra with an adjusted Elliot model for quantum-wells to the absorption of the NPLs in solution for bare, 1ML, 2ML and 3ML (a-d) respectively.** The yellow dashed line is the HH contribution, the green dashed line the LH contribution and the red line the sum of the two. The distinction from free-carrier absorption (broadened errorfunction) and exciton contribution (Lorentzian) can be clearly seen. The range of the data used for fitting is displayed by the plotted range of the fits data itself.

## Section S3 – Electrochemical charging of NPL films.

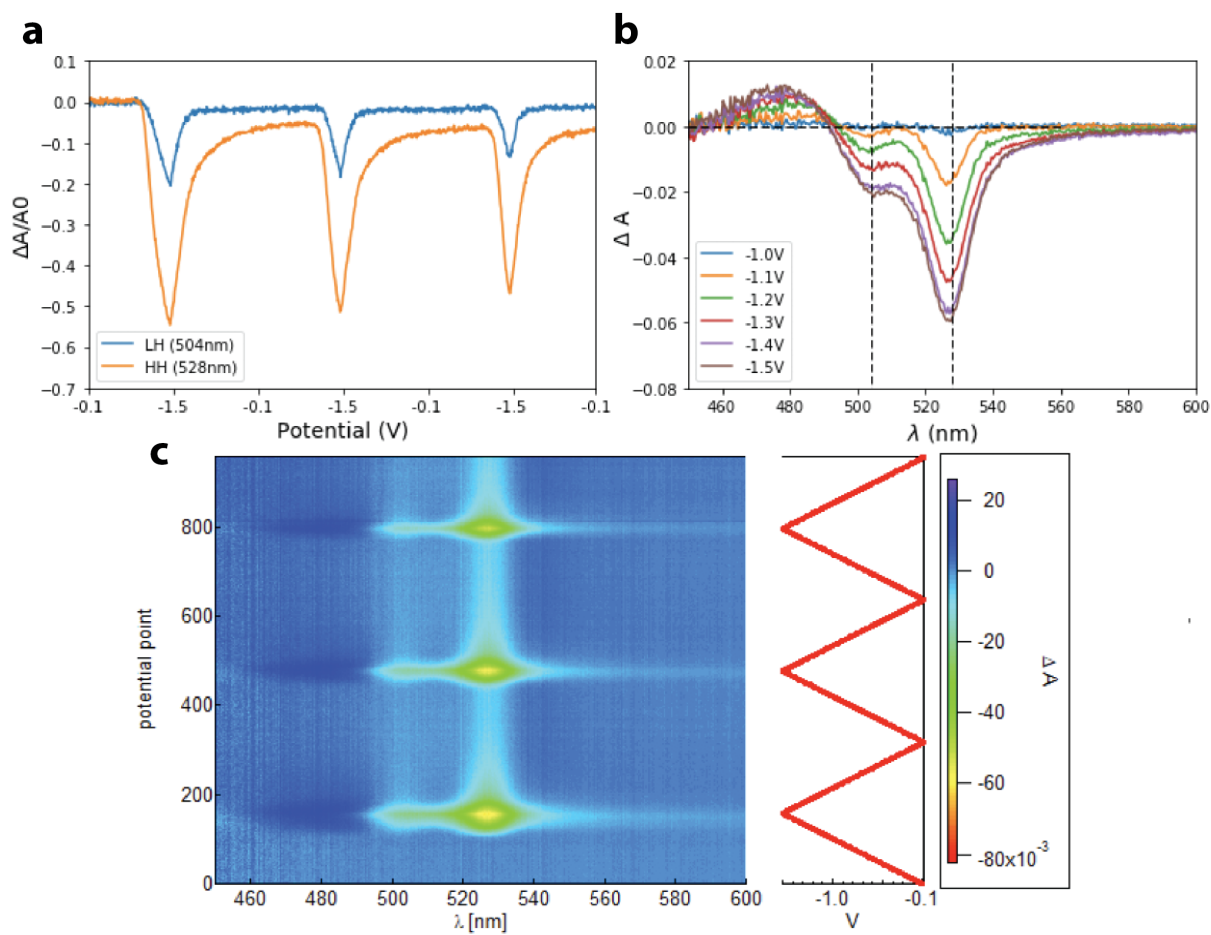

**Figure S6: Spectroelectrochemistry on a film of core-only CdSe NPLs.** We did not observe any photoluminescence from the film. **(a)** Fractional bleach (DA/A$_0$) versus applied electrochemical potential for the HH and LH transitions. We did not obtain a full bleach (DA/A$_0$). The LH bleach is likely convoluted with a shifted feature. **(b)** Spectral cuts at different applied electrochemical potentials. **(c)** Two-dimensional absorbance SEC map. Charge extraction is much slower than charge injection; the bleach features disappeared after the measurement, when we forced the system to stay at the open-circuit potential for 30 minutes. The sub-bandgap 'bleach' feature is likely a change in reflectivity of the film.

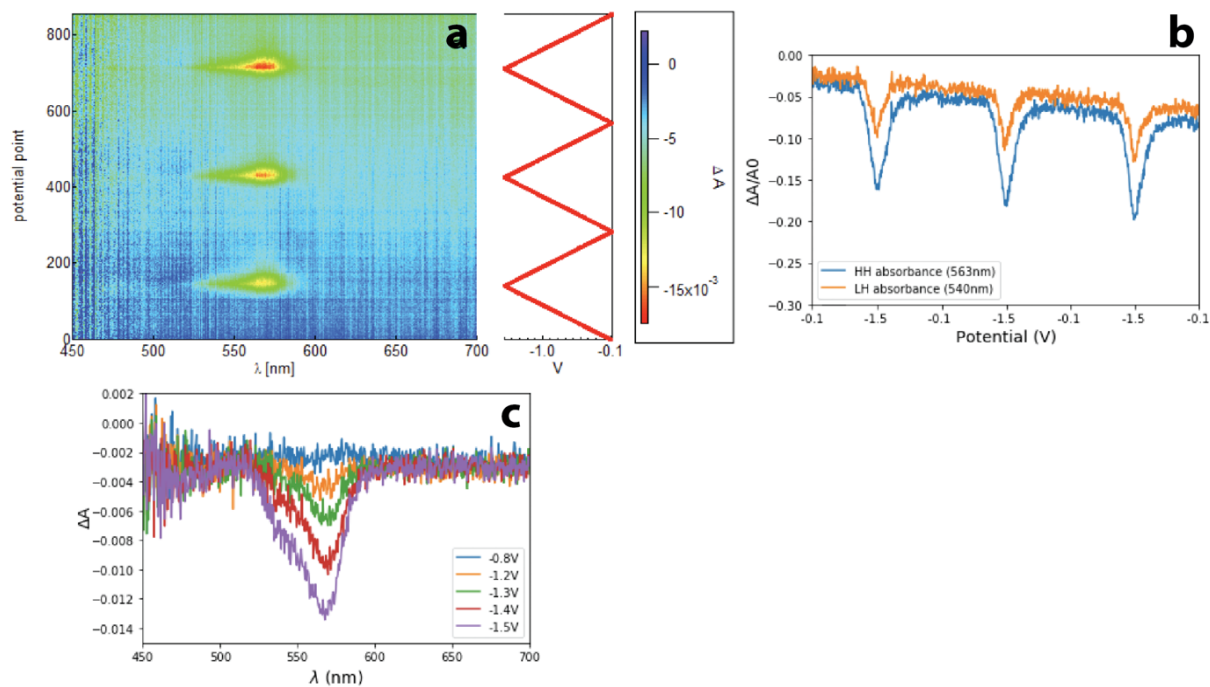

**Figure S7: Spectroelectrochemistry on a film of core-shell CdSe/1CdS NPLs.** We did not observe any photoluminescence from the film. **(a)** Two-dimensional absorbance SEC map. **(b)** Fractional bleach vs. applied electrochemical potential for the LH and HH transitions. **(c)** DA spectra at different applied electrochemical potentials.

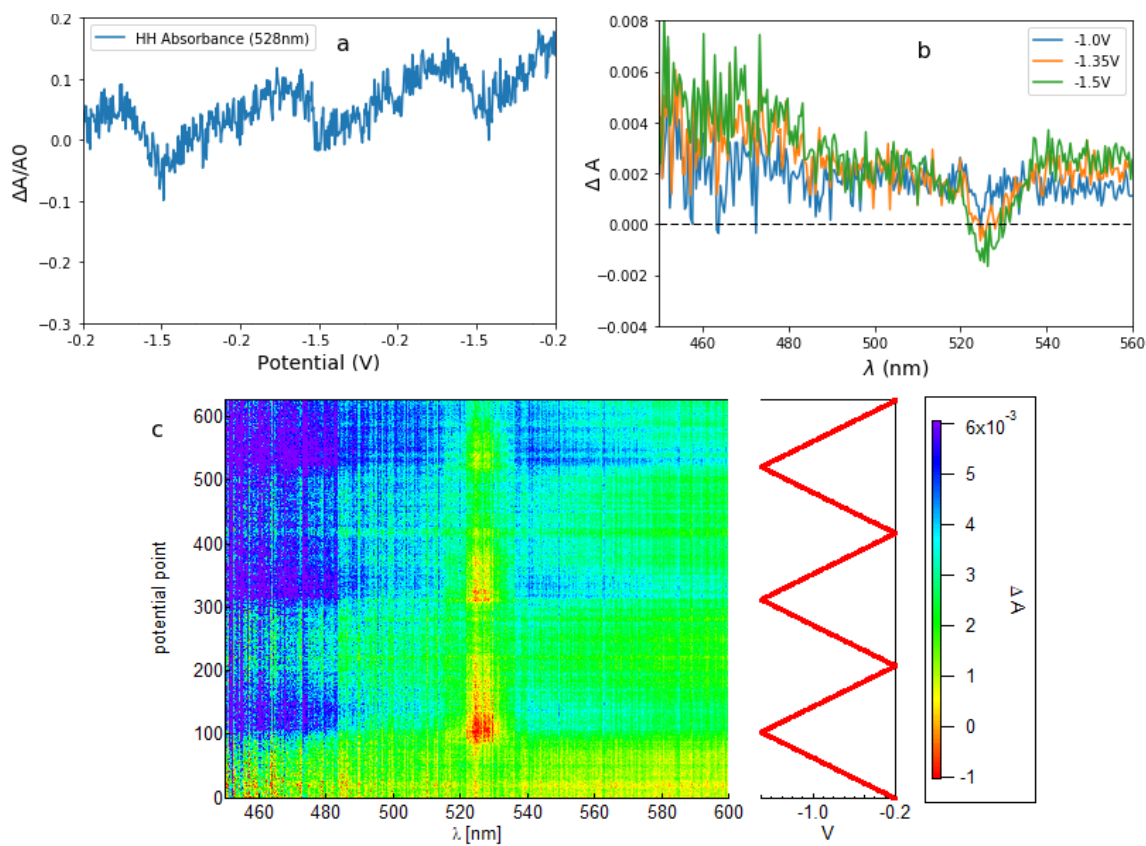

**Figure SXX: Spectroelectrochemistry on a film of core-crown CdSe/CdS NPLs.** We did not observe any photoluminescence from the film. **(a)** Fractional bleach as a function of applied potential. **(b)** DA spectra at different applied electrochemical potentials. **(c)** Two-dimensional absorbance SEC map.

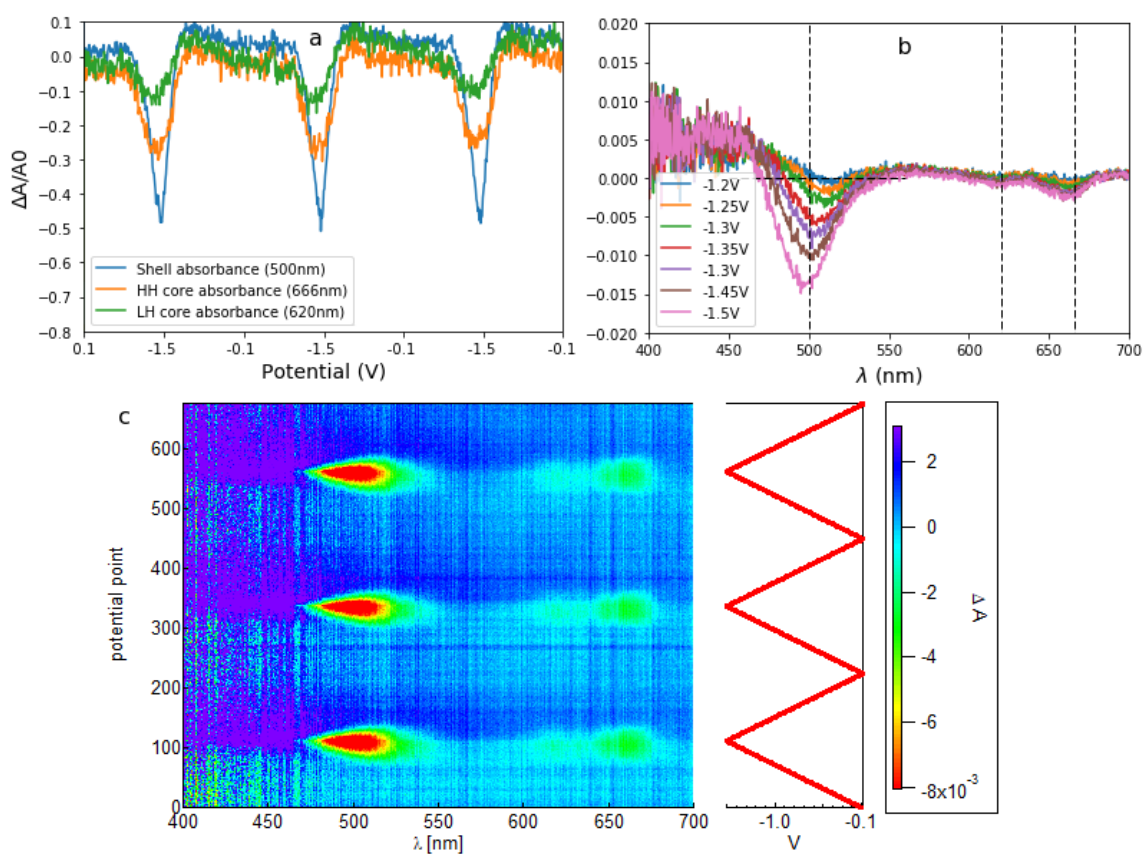

**Figure S8: Representative absorption spectroelectrochemistry on a film of core-shell CdSe/6CdS NPLs. (a)** Fractional bleach as a function of applied potential. **(b)** DA spectra at different applied electrochemical potentials. **(c)** Two-dimensional absorbance SEC map. Note that thinner CdS shells led to less reversible charging and discharging of the NPL films.

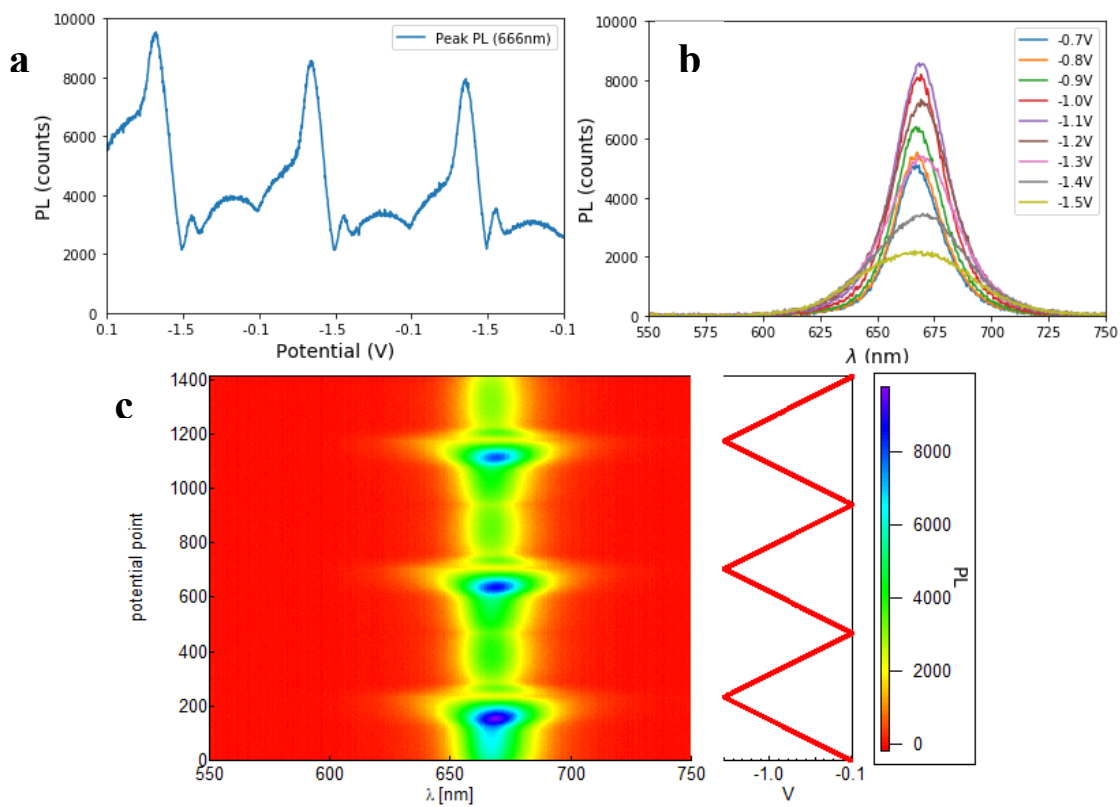

**Figure S9: Representative PL spectroelectrochemistry on a film of core-shell CdSe/6CdS NPLs.** **(a)** Integrated photoluminescence intensity as a function of applied potential. **(b)** PL spectra at different applied electrochemical potentials. **(c)** Two-dimensional PL SEC map. Note that thinner CdS shells lead to less reversible charging and discharging of the NPL films.

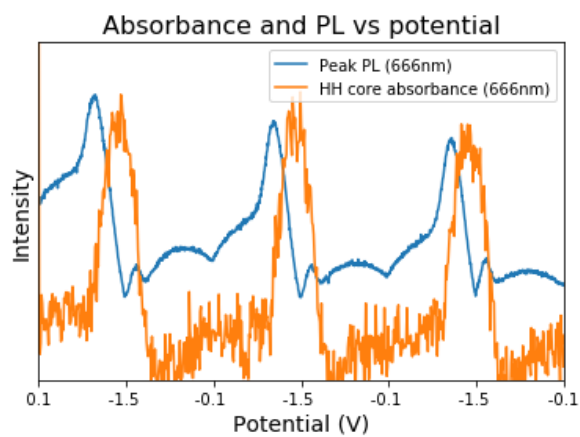

**Figure S10: Comparison between absorption bleach onset and PL quenching versus applied electrochemical potential in the CdSe/6CdS core-shell NPL film.**

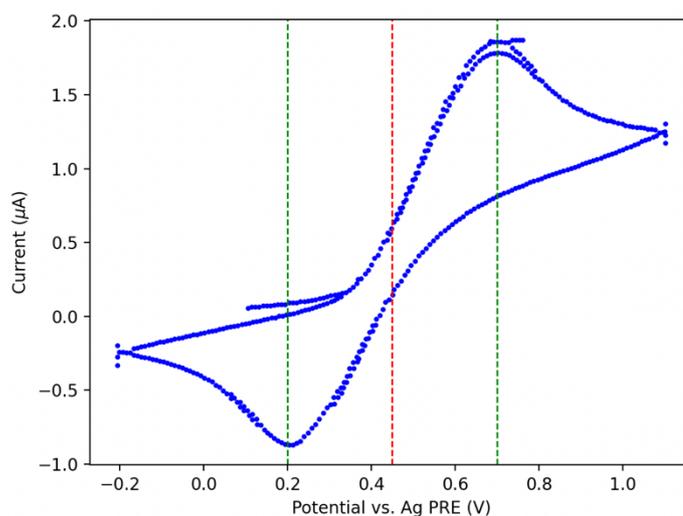

**Figure S11: Calibration of the Ag pseudo-reference electrode with the Fc/Fc$^+$ redox couple before the electrochemical TA experiments.** The green lines indicate the positions of the oxidation and reduction waves (+0.7 V and +0.2V respectively), and its' potential is given by the red line at +0.45 V. Since Fc/Fc$^+$ lies -4.7 eV below vacuum, the Ag PRE lies at -4.35 eV (or + 4.35 V) below vacuum. No significant shift of the reference potential was found after the SEC-TA experiments.

Section S4 – Temperature dependent absorption and PL of the CSS NPL film.

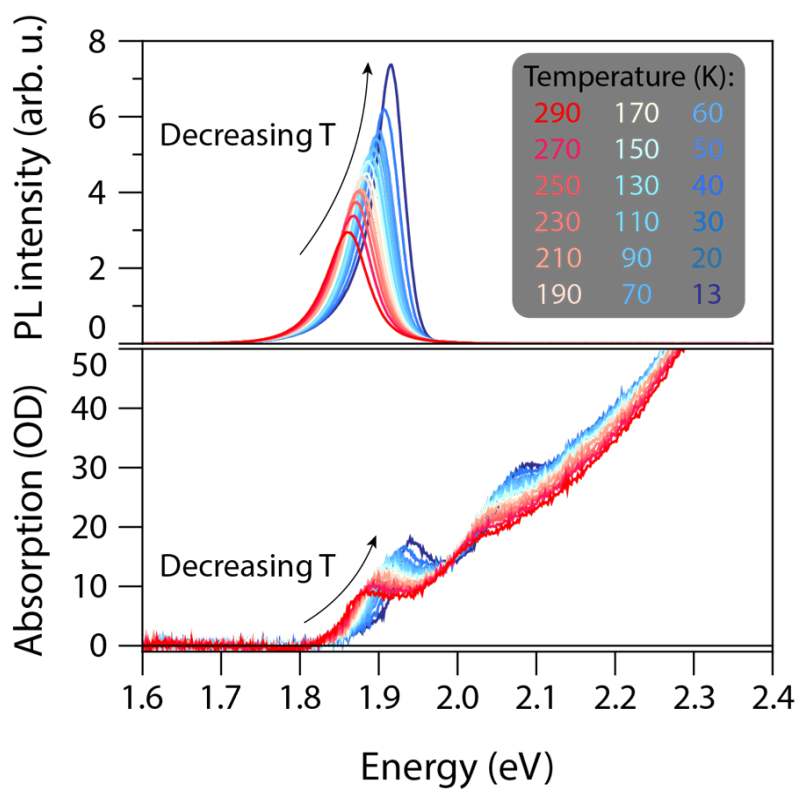

**Figure S12: Temperature dependent absorption and photoluminescence of a film of core-shell-shell NPLs.**

## Section S5 - NPL absorption cross-section determination

We determine the NPL absorption cross section using three methods:

1. **Poisson statistics of Auger recombination.** The magnitude of the bleach between 1-3 ns pump-probe delay time is smaller than directly after photoexcitation, but nonzero, meaning that there still is a finite population of nanocrystals which have excitons in there. The amplitude of the absorption bleach scales with the number of excitons present, and can be estimated via;

$$|\Delta A_{1-3\,ns}| \propto 1 - P_0 = 1 - e^{-\langle N \rangle}$$

   Where $|\Delta A_{1-3ns}|$ is the magnitude of the bleach between 1-3 nanoseconds, $P_0$ is the Poisson probability of finding zero excitons and is the average exciton population per nanocrystal. In turn, $\langle N \rangle = \sigma J_0$, with $\sigma$ being the absorption cross section at the excitation wavelength, and $J_0$ the incoming photon fluence. By fitting the data to the above equation, we obtain an absorption cross section at 400 nm of $5.6 \pm 0.2 \cdot 10^{-14}$ cm$^2$, which we use to calculate per photon fluence used in the TA experiments. For this analysis, we also correct the incoming photon fluence $J_0$ for absorption throughout the solution:

$$J'_0 = \frac{1 - e^{-\alpha L}}{\alpha L} J_0$$

   with the average of the photon fluence across the solution $J'_0$ length and $\alpha$ the absorption coefficient at the excitation wavelength. The term $\alpha \cdot L$ equals $A \cdot \ln(10)$, with A the optical density at 400 nm.

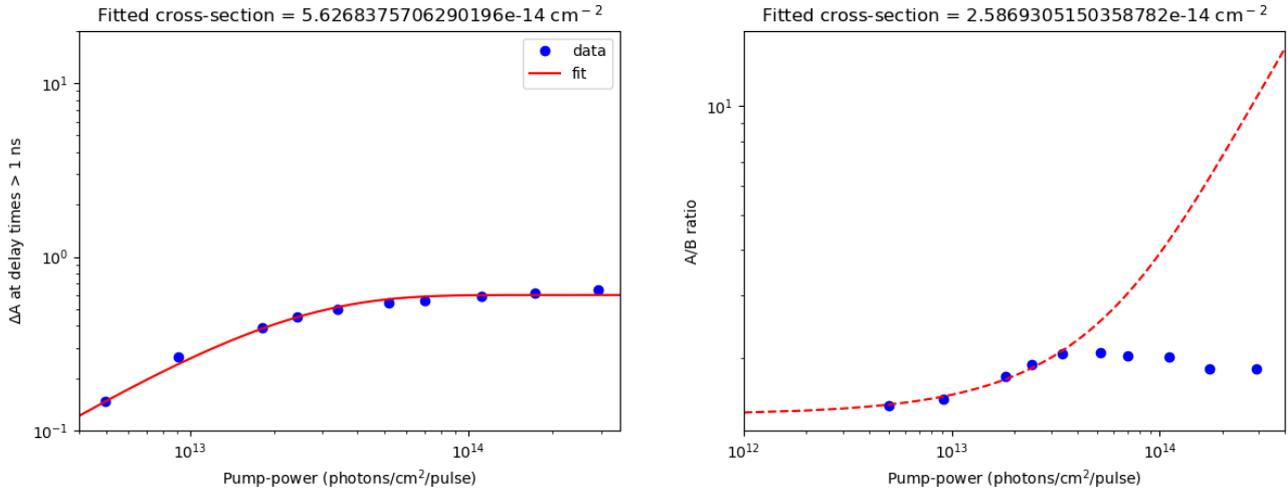

**Figure S13: NPL cross section determination using Poisson statistics of Auger recombination (left) and the 'A-over-B-ratio' (right).** The data in the left panel was fitted with equation SXX and resulted in an absorption cross-section at 400 nm of $5.6\pm0.2\cdot10^{-14}$ cm$^2$. The data in the right panel was fitted with equation SXX – over the first 5 datapoints - and resulted in an absorption-cross section at 400 nm of $2.6\pm0.4\cdot10^{-14}$ cm$^2$.

2. **'A-over-B ratio'.** The A/B ratio is used frequently in carrier multiplication experiments to determine quantum yields. The A-over-B ratio is defined as:

$$\frac{A}{B} = \frac{\Delta A_{t=0}}{\Delta A_{1-3ns}} = \frac{\sigma J'_0}{1-e^{-\sigma J'_0}}$$

We determined the A/B ratio for various fluences, and fit with the above equation to obtain a cross section of $2.6\pm0.4\cdot10^{-14}$ cm$^2$. Note that we only fit the initial 5 datapoints, since the equation does not take into account the saturation of the early-time bleach at higher fluences. The result is shown in Figure SXX. The cross section determined with this method is 2.2 times lower than determined via method 1.

3. **Calculation of the absorption cross-section based on known dielectric constants.**

   We estimate the absorption cross section of our CSS NPLs based on known dielectric functions of the bulk materials, obtained from https://refractiveindex.info/. We find values for $k$ at 320 nm.

   CdSe: Ninomiya and Adachi 1995: Cubic. $k = 1.1052$.
   CdS: Ninomiya and Adachi 1995: $k = 0.65143$.
   ZnS: Ozaki and Adachi 1994: Cubic ZnS. $k = 0.3484$.

   From $k$ we calculate the intrinsic absorption coefficient $\mu$, as $\mu = \frac{4\pi k}{\lambda}$. Using the number of CdS (5) and ZnS (2) monolayers we predict from the synthesis on top of our NPLs, we calculate the volume weighted intrinsic absorption coefficient (weighing each intrinsic absorption coefficient by the volume of that material present), and obtain a total intrinsic absorption coefficient at 320 nm of $1.85 \cdot 10^5 \text{cm}^{-1}$. From the volume of the CSS NPLs $V_{NPL}$ (1418.1 nm$^3$), we calculate the absorption cross section at our pump wavelength (400 nm) as

   $$\sigma_{400\ nm} = \mu_{320\ nm}\ V_{NPL}\ \frac{A_{400\ nm}}{A_{320\ nm}}$$

   where we use the ratio between the absorption at 400 nm and 320 nm to obtain the absorption cross section at our pump wavelength – $9.4 \cdot 10^{-14}\ \text{cm}^2$. This value is 1.7 times higher than the value obtained via method 1.

   Since method 2 produces a lower and method 3 produces a higher cross section, we decide to use the cross section from method 1 throughout the paper, which is roughly the average from all three methods.

Section S6 – Particle-in-a-box calculations of the confinement energies and thresholds for optical gain.

For a particle in a 2D box, the confinement energy is given by:

$$E_{confinement} = \frac{h^2}{8m^*}\left(\left[\frac{n_x}{L_x}\right]^2 + \left[\frac{n_y}{L_y}\right]^2 + \left[\frac{n_z}{L_z}\right]^2\right)$$

With $h$ being Planck's constant, $m^*$ the effective mass of the particle (0.13 for the electrons and 0.3 for holes in CdSe), $n_{x,y,z}$ the quantum numbers in the x, y and z directions and $L_{x,y,z}$, the lengths of the box in the x, y, and z directions. For the CSS NPLs we used ($L_x$ = 24.9 nm, $L_y$ = 9.9 nm), we calculate a total confinement energy in the NPL plane (we ignore the thickness), i.e. the sum of the electron and hole confinement energies, of 49 meV. For the difference in energy from the first to second quantized level in the longest direction ($n_x$, $n_y$ = 1,1 to $n_x$, $n_y$ = 2,1) we find an energy of 20 meV.

**Threshold for optical gain**

For a 2D bulk sheet of CdSe, there should be a continuous density-of-states in the x and y directions. We calculate the threshold for optical gain by calculating at which concentration of electrons and holes the quasi-Fermi level is equal to the conduction band and valence band, respectively. For bulk semiconductors this is standard, since for all levels below the quasi-Fermi level the occupation is larger than 0.5, and hence there is population inversion.

For the 2D sheets, we calculate the energy levels based on a simple particle-in-a-box calculation, again without excitonic effects. We again calculate the density at which the quasi-Fermi levels are equal to the first empty and filled states (i.e. the LUMO and HOMO levels in the NPLs).

**Modelling optical gain in NPLs: optimal NPL lateral size for lowest threshold excitation fluence**

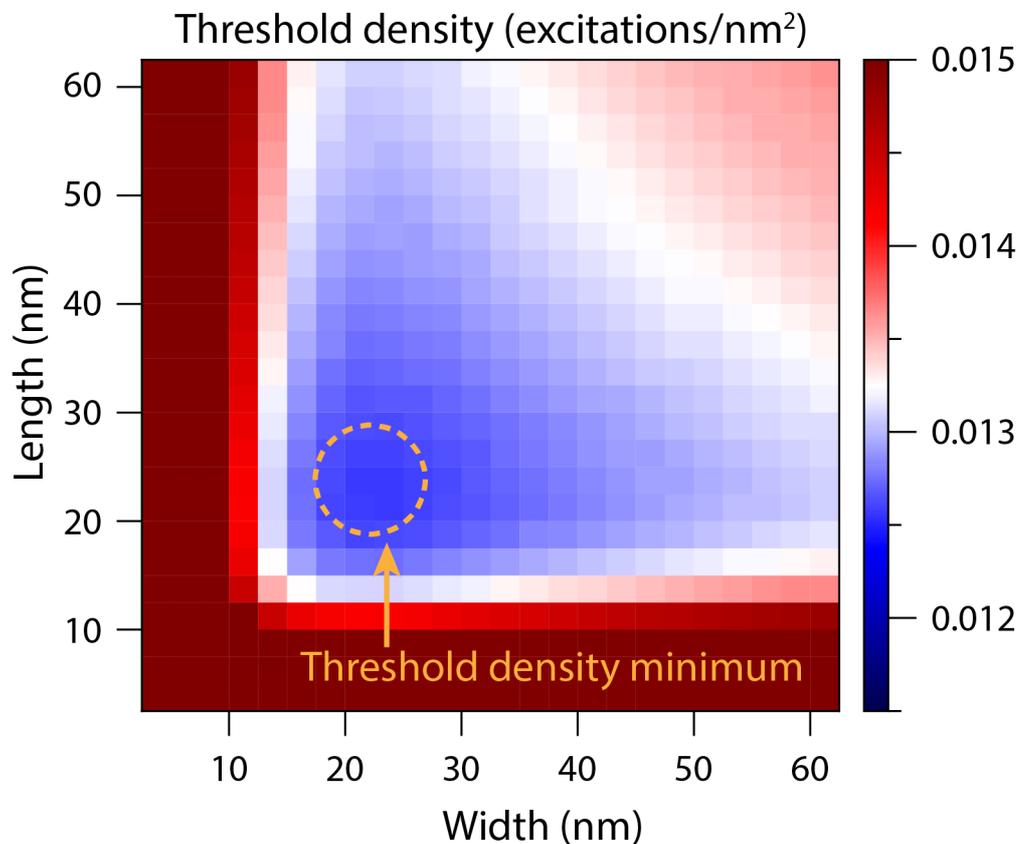

**Figure S14: Modelling the optical gain in NPLs with a particle-in-a-box model.** Threshold density for optical gain from a state-filling model. There is an optimum in threshold density for a NPL lateral size of roughly 23 by 23 nm.

Section S7 - Heisenberg model and state-filling in NPLs

We find that we can fully bleach the lowest excitonic transition in CdSe/CdSe/ZnS NPLs by electrochemically injecting electrons. We can also fully bleach the transition by optical excitation. However, there is a large difference in TA signals for neutral and charged NPLs: neutral NPLs show sharp bleach features, charged ones show extreme broadening.

The question is how to explain the absorption bleach features (both electrochemical and optical): do they result from state filling or from screening of the e-h Coulomb interaction (Mott transition). Our tentative conclusion is that at low excitation densities (<N> smaller than 1) we can explain most things by state filling alone, but we cannot exclude <u>both effects happen at higher excitation densities</u>. Screening results in a strong decrease of the exciton binding energy, resulting in fast exciton dissociation and corresponding lifetime broadening. This is especially important for the electrochemically charged case. The observed broadening corresponds to an exciton (or rather, charged exciton) lifetime of ~40 fs.

This does not mean state filling is not important for electrochemical charging. The question is how to model that, since we fill up electron states, but we observe a bleach of the exciton transition. Since excitons and electrons are described by different wavefunctions, the exciton wavefunction should be expended in electron states, or vice versa. Here we use a rather hand waving estimate based on the Heisenberg uncertainty principle to estimate up to what energy (or up to what Fermi-level) the $n_z = 1$ band needs to be filled to fully bleach the 1S exciton.

**Model based on Heisenberg's uncertainty relation**

Starting from Heisenberg:

$$\Delta x \, \Delta k \geq \frac{1}{2}$$

Where $\Delta x$ and $\Delta k$ are the (Gaussian) uncertainties in position and momentum. For the uncertainty in position, we make an assumption that this is roughly the exciton Bohr radius: $\Delta x \approx a_B$, changing the uncertainty relation into $a_B \, \Delta k \geq \frac{1}{2}$.

Now we can express the Bohr radius in a material by the Bohr radius of the 1S hydrogen wavefunction $a_B = \frac{a_0 \varepsilon_R}{\mu}$, where $\varepsilon_R$ is the real part of the static dielectric constant and $\mu$ the reduced effective mass, which can be expressed as

$$\frac{1}{\mu} = \frac{1}{m_e^*} + \frac{1}{m_h^*} \approx \frac{1}{m_e^*}$$

Here we use that the effective mass of the electron (0.13) is three times lower than that of the hole (0.3) in zinc-blende CdSe.

Furthermore, we can express in terms of a set of constants;

$$a_0 = \frac{4\pi \varepsilon_0 \hbar^2}{m_0 e^2}$$

We can write an expression for the uncertainty of the total energy in terms of the uncertainty in in-plane wavenumber and

$$\Delta E_{tot} = \frac{\hbar^2 \Delta k_x^2}{2 \, m_e^* \, m_0} + \frac{\hbar^2 \Delta k_y^2}{2 \, m_e^* \, m_0} = \frac{\hbar^2 \Delta k^2}{m_e^* \, m_0}$$

Using the above equations, we find that

$$\Delta E_{tot} = \frac{\hbar^2 \Delta k^2}{m_e^* \, m_0} = \frac{\hbar^2}{4 \, m_e^* \, m_0 \, a_B^2} = \frac{\hbar^2 \, m_e^{*2}}{4 \, m_e^* \, m_0 \, a_0^2 \, \varepsilon_R^2} = \frac{\hbar^2 \, m_e^*}{4 \, m_0 \, a_0^2 \, \varepsilon_R^2}$$

When we use accepted values for the bulk exciton Bohr radius in CdSe, 5.4 nm, an effective mass of the electron of 0.13, and the static dielectric constant in CdSe of 10.2, we for $\Delta E_{tot}$ of about 31 meV.

**Discussion**

We calculated a value of roughly 31 meV from our model, compared to the roughly 140 mV necessary to fully bleach the HH exciton transition. This would suggest that state filling should more quickly bleach the exciton transition than observed.

A reason why this value is too small, is that the dielectric constant should be roughly 2 times smaller in a two-dimensional system. This would make the estimated energy ~40 meV. Even when we vary this around a bit, we still get an estimate that is smaller than the 140 mV we measure experimentally.

Furthermore, as discussed throughout the main text, we also mention that the obtained width of the bleach vs. potential curve can be larger due to drops of the electrochemical potential (change in potential does not correspond 1:1 to change in Fermi level in the NPL film).

## Section S8 - Additional discussion based on Schmidt-Rink et al. [5]

Optical excitation or electronic doping leads to the occupation of some free carrier states that are now no longer available for the formation of new excitons. This leads to a decrease of the exciton absorption, an effect that is called state filling. In the case of a large exciton binding energy, and a correspondingly small Bohr radius, the number of k-states that contributes to the exciton wavefunction is large, and the bleach of a single free carrier (occupying a state near k=0) is small. For a smaller binding energy this bleach will be larger.

Considering only the effect of state-filling on the absorption bleach Schmitt-Rink et al. derived that the saturation density $N_S$, the excitation density where the exciton absorption is fully bleached, is equal to[5]

$$\frac{1}{N_S} = 8\pi a_{X,2D}^2 \tag{a1}$$

Eq. a1 is valid, according to Schmitt-Rink, for above resonant excitation (as we use in our experiments), which results directly in the formation of free electrons and holes rather than excitons and for the case when kT << $E_{B,HH}$. Entering the exciton Bohr radius of 3.2 nm, derived from the experimentally determined exciton binding energy, this gives a saturation density of $N_S = 3.9 \cdot 10^{11}$ cm$^{-2}$, *exactly* in line with the experimentally determined gain threshold. For excitons (e.g. formed by resonant excitation and before exciton dissociation takes place), the expected saturation density is given by[5]

$$\frac{1}{N_S} = \frac{32}{7}\pi a_{2D}^2 \tag{a2}$$

which gives a saturation density of $N_S = 6.8 \cdot 10^{11}$ cm$^{-2}$. The saturation density for free-carrier excitation matches our experimentally observed saturation density better than the saturation density for excitons. Note that this does not mean that the observed optical gain stems from free-carrier species, since the probe pulse in the experiment still generates excitons around 1.9 and 2.0 eV.

The above arguments based on state filling explain why the exciton bleaches in our NPLs, without invoking screening of Coulomb or exchange interactions. As the saturation density in our NPLs corresponds to 1 exciton/NPL, we cannot rule out that at higher excitation densities, Coulomb and exchange effects do play a crucial role. Indeed, severe broadening of the excited state absorption is observed in NPLs where the excitation density is high [$<N> \gg 1$, Figure 2(d) of the main text]. Again, this indicates that the core-shell-shell NPLs act as particle-in-a-box-like systems.

## REFERENCES

(1) Rossinelli, A. A.; Rojo, H.; Mule, A. S.; Aellen, M.; Cocina, A.; De Leo, E.; Schäublin, R.; Norris, D. J. Compositional Grading for Efficient and Narrowband Emission in CdSe-Based Core/Shell Nanoplatelets. *Chem. Mater.* **2019**, *31* (22), 9567–9578. https://doi.org/10.1021/acs.chemmater.9b04220.

(2) Kelestemur, Y.; Shynkarenko, Y.; Anni, M.; Yakunin, S.; De Giorgi, M. L.; Kovalenko, M. V. Colloidal CdSe Quantum Wells with Graded Shell Composition for Low-Threshold Amplified Spontaneous Emission and Highly Efficient Electroluminescence. *ACS Nano* **2019**, *13* (12), 13899–13909. https://doi.org/10.1021/acsnano.9b05313.

(3) Ithurria, S.; Talapin, D. V. Colloidal Atomic Layer Deposition (c-ALD) Using Self-Limiting Reactions at Nanocrystal Surface Coupled to Phase Transfer between Polar and Nonpolar Media. *J. Am. Chem. Soc.* **2012**, *134* (45), 18585–18590. https://doi.org/10.1021/ja308088d.

(4) Tomar, R.; Kulkarni, A.; Chen, K.; Singh, S.; Van Thourhout, D.; Hodgkiss, J. M.; Siebbeles, L. D. A.; Hens, Z.; Geiregat, P. Charge Carrier Cooling Bottleneck Opens Up Nonexcitonic Gain Mechanisms in Colloidal CdSe Quantum Wells. *J. Phys. Chem. C* **2019**, *123* (14), 9640–9650. https://doi.org/10.1021/acs.jpcc.9b02085.

(5) Schmitt-Rink, S.; Chemla, D. S.; Miller, D. A. B. Theory of Transient Excitonic Optical Nonlinearities in Semiconductor Quantum-Well Structures. *Phys. Rev. B* **1985**, *32* (10), 6601–6609. https://doi.org/10.1103/PhysRevB.32.6601.